\documentclass[preprint,12pt]{elsarticle}

\usepackage{amssymb}
\usepackage{multirow}
\usepackage{booktabs}
\usepackage{amssymb}
\usepackage{mathrsfs}
\usepackage{longtable}
\usepackage{lscape}
\usepackage{tabularx}
\usepackage{epsfig}
\usepackage{colortbl}
\usepackage{graphicx}
\usepackage{subfigure}
\definecolor{gray}{gray}{.9}

\journal{Physica A}

\begin{document}

\begin{frontmatter}



\title{Nonextensive analysis on the local structure entropy of complex networks}


\author[swu]{Qi Zhang}
\author[swu]{Meizhu Li}
\author[swu]{Yuxian Du}
\author[swu,NWPU,vu]{Yong Deng\corref{cor}}
\ead{ydeng@swu.edu.cn, prof.deng@hotmail.com}
\author[vu]{Sankaran Mahadevan}

\cortext[cor]{Corresponding author: Yong Deng, School of Computer and Information Science, Southwest University, Chongqing, 400715, China.}

\address[swu]{School of Computer and Information Science, Southwest University, Chongqing, 400715, China}
\address[NWPU]{School of Automation, Northwestern Polytechnical University, Xian, Shaanxi 710072, China}
\address[vu]{School of Engineering, Vanderbilt University, Nashville, TN, 37235, USA}

\begin{abstract}
The local structure entropy is a new method which is proposed to identify the influential nodes in the complex networks. In this paper a new form of the local structure entropy of the complex networks is proposed based on the Tsallis entropy. The value of the entropic index $q$ will influence the property of the local structure entropy. When the value of $q$ is equal to 0, the nonextensive local structure entropy is degenerated to a new form of the degree centrality. When the value of $q$ is equal to 1, the nonextensive local structure entropy is degenerated to the existing form of the local structure entropy. We also have find a nonextensive threshold value in the nonextensive local structure entropy. When the value of $q$ is bigger than the nonextensive threshold value, change the value of $q$ will has no influence on the property of the local structure entropy, and different complex networks have different nonextensive threshold value.

The results in this paper show that the new nonextensive local structure entropy is a generalised of the local structure entropy. It is more reasonable and useful than the existing one.

\end{abstract}
\end{frontmatter}

\section{Introduction}
\label{Introduction}
The complex networks is a model which can used to describe those complex relationship in the real system, such as the biological, social and technological systems \cite{albert2000error,newman2003structure}. Many property of the complex networks have illuminated by these researchers in this filed, such as the network topology and dynamics \cite{watts1998collective,newman2006structure}, the property of the network structure \cite{newman2003structure,barthelemy2004betweenness}, the self-similarity and fractal property of the complex networks\cite{song2005self,wei2014informationdimension}, the controllability and the synchronization of the complex networks \cite{liu2011controllability,arenas2008synchronization} and so on \cite{barrat2004architecture,barabasi2009scale,barabasi1999emergence,barabasi2009scale,song2005self,teixeira2010complex}.
How to identify the influential nodes in the complex networks has attracted many researchers to study it. Recently, a local structure entropy of the complex networks is proposed to identify the influential nodes in the complex networks \cite{zhang2014local}. In the local structure entropy the node's influence on the whole network is replaced by the local network. The degree entropy of the local network is used as the measure of the influence of the node on the whole network. The local structure entropy is based on the shannon entropy. In this paper the Tsalli entropy which is proposed by Tsalli et.al \cite{tsallis1988possible} is used to analysis the property of the local structure entropy.
Depends on the nonextensive statistical mechanics, the relationship between each node can be described by the nonextensive additivity. 

In this paper, the property of the local structure entropy is analysed by the nonextensive statistical mechanics.
Depends on the Tsallis entropy, a new form of the local structure entropy is proposed in this paper. In the nonextensive local structure entropy, the influences of the node on the whole network is changed by the entropic index $q$. The nonextensive in the local structure entropy is changed correspond to value of $q$ and the nonextensive additivity is restricted by the value of $q$. We also find the nonextensive threshold value of $q$ in the nonextensive local structure entropy. When the value of $q$ is bigger than the nonextensive threshold value, then the property of the local structure entropy will not be controlled by the $q$. When the value of $q$ is equal to 0, then the local structure entropy is degenerated to another form of the degree centrality. The nonextensive local structure entropy is a generalised of the local structure entropy.

The rest of this paper is organised as follows. Section \ref{Rreparatorywork} introduces some preliminaries of this work, such as the local structure entropy of complex networks and the nonextensive statistical mechanics. In section \ref{newmethod}, the analysis of the local structure entropy based on the nonextensive is proposed. The application of the nonextensive analysis in these real networks is shown in the section \ref{application}. Conclusion is given in Section \ref{conclusion}.
\section{Preliminaries}
\label{Rreparatorywork}

\subsection{Local structure entropy of complex networks}
\label{Existing methods}
There many methods are proposed to identify the influential nodes in the complex networks. The degree centrality and the betweenneess centrality are the wildly used method to identify the influential nodes in the complex networks.
Recently, the "Local structure entropy" of the complex networks which is based on the degree centrality and the shannon entropy is proposed \cite{zhang2014local}. The details of the local structure entropy of the complex networks is shown as follows \cite{zhang2014local}.

The definition of the local structure entropy can be divided into three steps, the details are shown as follows \cite{zhang2014local}.

\textbf{Step 1 Creating a local network}: First, choose one of the node in the network as the central node. Second, find all of the nodes in the network which are connect with the central node in directly. Third, create a local network which contains the central node and his neighbour nodes.

\textbf{Step 2 Calculating the unit of the local structure entropy}: Calculate the degree of each node and the total number of the degree in the local network. The unit of the local structure entropy can be represents as the ${p_{ij}}$, it is defined in the Eq.(\ref{P_ij}).

\textbf{Step 3 Calculating the local structure entropy of each node}: The definition of the local structure entropy for each node is shown in the Eq.(\ref{Local_E}).

The definition of the local structure entropy of the complex networks is shown as follows \cite{zhang2014local}.

 \begin{equation}\label{Local_E}
    {LE_i} = -\sum\limits_{j = 1}^n {{p_{ij}}} \log {p_{ij}}
 \end{equation}

Where the ${LE_i}$ represents the local structure entropy of the $i$th node in the complex networks. The $n$ is the total number of the nodes in the local network. The ${p_{ij}}$ represents the percentage of degree for the $j$th node in the local network. The definition of the ${p_{ij}}$ is shown in the Eq.(\ref{P_ij}).

\begin{equation}\label{P_ij}
{p_{ij}} = \frac{{\ degree(j)}}{{\sum\limits_{j = 1}^n {\ degree(j)} }}
\end{equation}

\begin{figure}[!htbp]
    \centering

    \subfigure[The example network A ]{
    \label{karate_E:subfig:a} 
    \centering
    \includegraphics[scale=0.6]{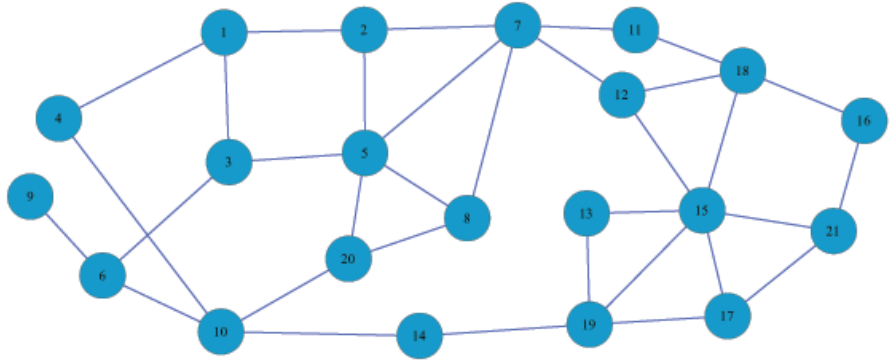}}
  \subfigure[The 5$th$ node. ]{
    \label{karate_E:subfig:a} 
    \centering
    \includegraphics[scale=0.8]{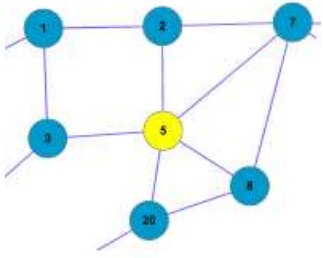}}
  \hspace{2cm}
    \subfigure[The 15$th$ node]{
    \label{karate_E:subfig:b} 
    \includegraphics[scale=0.8]{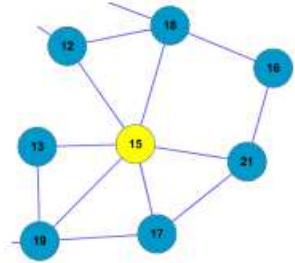}}
    \centering
  \caption{In the example network A, different node has different value of degree. The ${LE_i}$ of each node is different to each others. We use the node 5 and node 15 to show the details of the calculation of the local structure entropy of each node. First, the local structure entropy of the node 5. The node 5 has 5 neighbours, the node 2, 3, 20, 8, 7. The degree of each neighbour node is 3, 3, 3,  3, 5. The degree of the node 5 is 5. The total degree in the local network is 19. Then the set of the degree in the local network is ${D_5}={3/19, 3/19, 3/19, 5/19, 5/19}$. Then the $LE_5$=1.8684. Second, the local structure entropy of the node 15. The node 15 has 6 neighbours, the node 12, 13, 17, 18, 19, 21. The degree of each neighbour node is 3, 2, 3, 4, 4, 3. The degree of node 15 is 6. The total degree in the local network is 25. The set of the degree in the local network is ${D_15}={3/25, 2/25, 3/25, 4/25, 4/25, 3/25, 6/25}$ Then the $LE_15$=1.89426. From the definition of the local structure entropy the node 15 is more influential than the node 5 in the example network.}\label{E_E}
\end{figure}

An example of the process to calculate the local structure entropy is shown in the Fig.\ref{E_E}.

It is clear that in the local structure entropy the influence of the node on the whole network is replaced by the influence of the local network on the whole network \cite{zhang2014local}.
\subsection{Nonextensive statistical mechanics}
\label{Tsallis entropy}
The entropy is defined by Clausius for thermodynamics \cite{clausius1867mechanical}. For a finite discrete set of probabilities the definition of the Boltzmann-Gibbs entropy \cite{gibbs2010elementary} is given as follows:

\begin{equation}\label{S_BG}
{S_{BG}} =  -k \sum\limits_{i = 1}^N {{p_i}} \ln {p_i}
\end{equation}

The conventional constant $k$ is the Boltzmann universal constant for thermodynamic systems. The value of $k$ will be taken to be unity in information theory \cite{shannon2001mathematical}.

In 1988, a generalised entropy have been proposed by Tsallis  \cite{tsallis1988possible}. It is shown as follows:

\begin{equation}\label{S_q}
{S_q} =  - k\sum\limits_{i = 1}^N {{p_i}} {\ln _q}\frac{1}{{{p_i}}}
\end{equation}

The $q-logarithmic$ function in the Eq. (\ref{S_q}) is presented as follows:
\begin{equation}\label{ln_q}
{\ln _q}{p_i} = \frac{{{p_i}^{1 - q} - 1}}{{1 - q}}({p_i} > 0;q \in \Re ;l{n_1}{p_i} = ln{p_i})
\end{equation}

Based on the Eq. (\ref{ln_q}), the Eq. (\ref{S_q}) can be rewritten as follows:

\begin{equation}\label{S_q1}
{S_q} =  - k\sum\limits_{i = 1}^N {{p_i}} \frac{{{p_i}^{q - 1} - 1}}{{1 - q}}
\end{equation}

\begin{equation}\label{S_q2}
{S_q} =  - k\sum\limits_{i = 1}^N {\frac{{{p_i}^q - {p_i}}}{{1 - q}}}
\end{equation}

\begin{equation}\label{S_q1}
{S_q} = k\frac{{1 - \sum\limits_{i = 1}^N {{p_i}^q} }}{{q - 1}}
\end{equation}

Where ${N}$ is the number of the subsystems.

Based on the Tsallis entropy, the nonextensive theory is established by Tsallis et.al. The nonextensive statistical mechanics is a generalised statistical mechanics.
\section{Nonextensive analysis of the local structure entropy of complex networks}
\label{newmethod}

The main idea of the local structure entropy is try to use the influence of the local network to replace the influence of the node on the whole network \cite{zhang2014local}. However, in the definition of the local structure entropy of each node, the relationship between each node in the local network is extensive. In order to illuminate the property of the local structure entropy, in this paper the nonextensive statistical mechanic is used in the definition of the local structure entropy.

Depends on the Tsallis entropy, the definition of the local structure entropy is redefined as follows:

\begin{equation}\label{T_local_E}
{S_q}_i =  - k\sum\limits_{i = 1}^N {{p_i}} {\ln _q}\frac{1}{{{p_i}}}
\end{equation}
Where in the Eq.(\ref{T_local_E}), the logarithmic function in the local structure entropy is replaced by the $q-logarithmic$ function in the Eq.(\ref{ln_q}).
The ${{S_q}_i}$ is the new local structure entropy of the node $i$. It is defined based on the Tsallis entropy \cite{tsallis2009introduction}. The ${{p_i}_j}$ is defined in the Eq.(\ref{P_ij}). The $q$ is the nonextensive entropic index.

We use the example network A which is shown in the Fig.\ref{E_E} to show the nonextensive property of the local structure entropy.

\begin{figure}
    \centering

    \subfigure[q=0.1]{
    \label{Yeast-infect-local1} 
    \centering
    \includegraphics[scale=0.25]{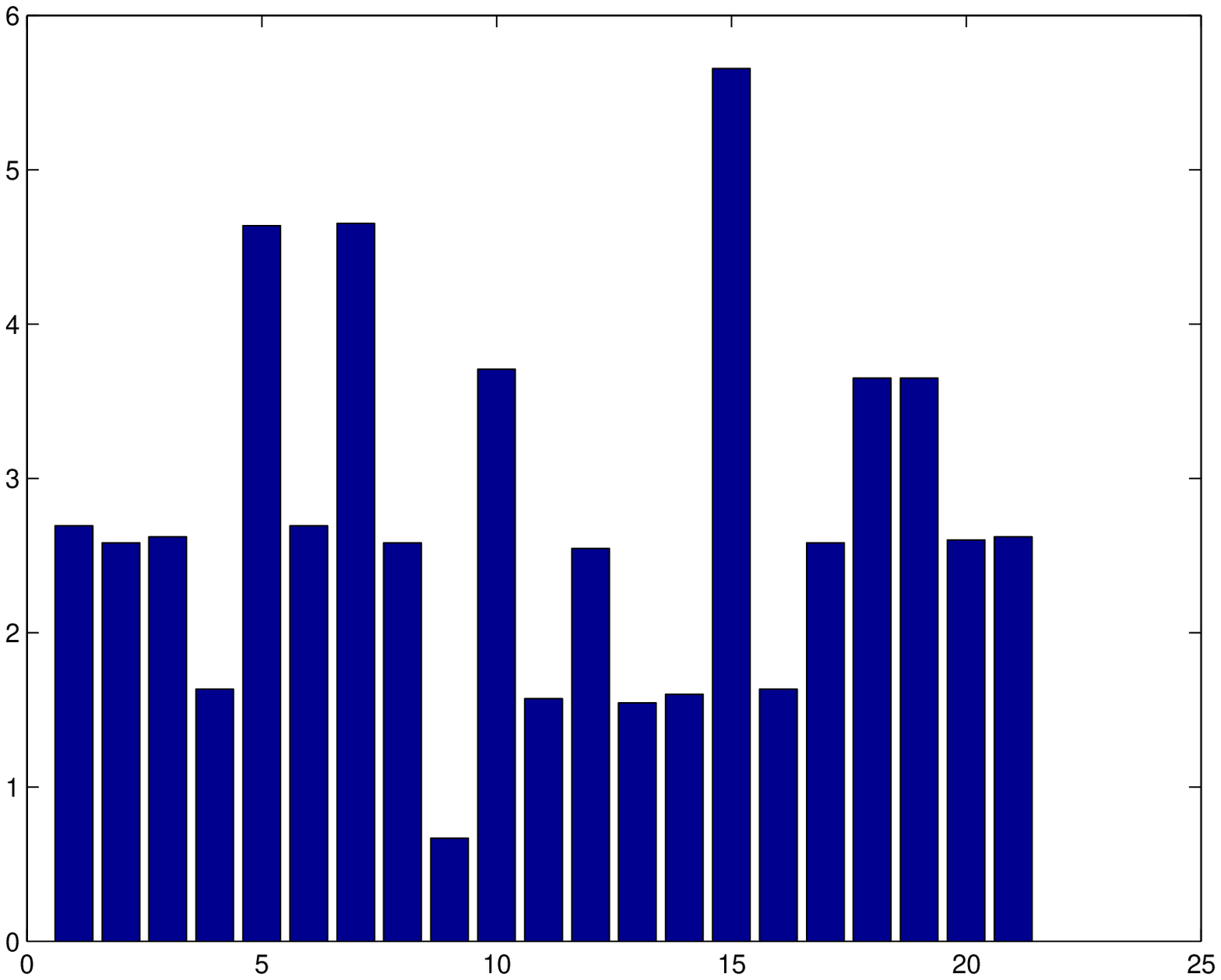}}
    \subfigure[q=0.2]{
    \label{Yeast-infect-local:a} 
    \centering
    \includegraphics[scale=0.25]{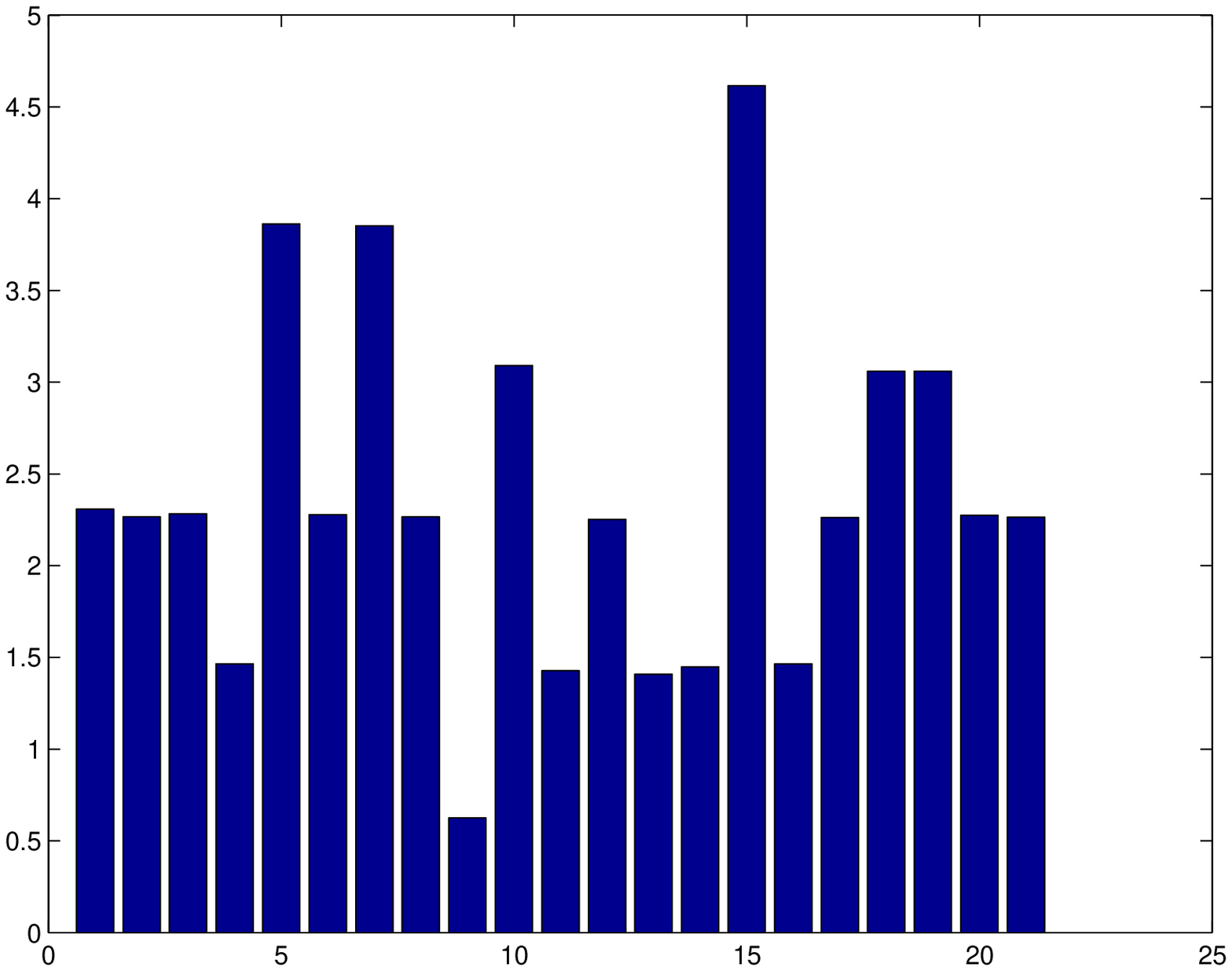}}
    \subfigure[q=0.3]{
    \label{Yeast-infect-local:b} 
    \centering
    \includegraphics[scale=0.25]{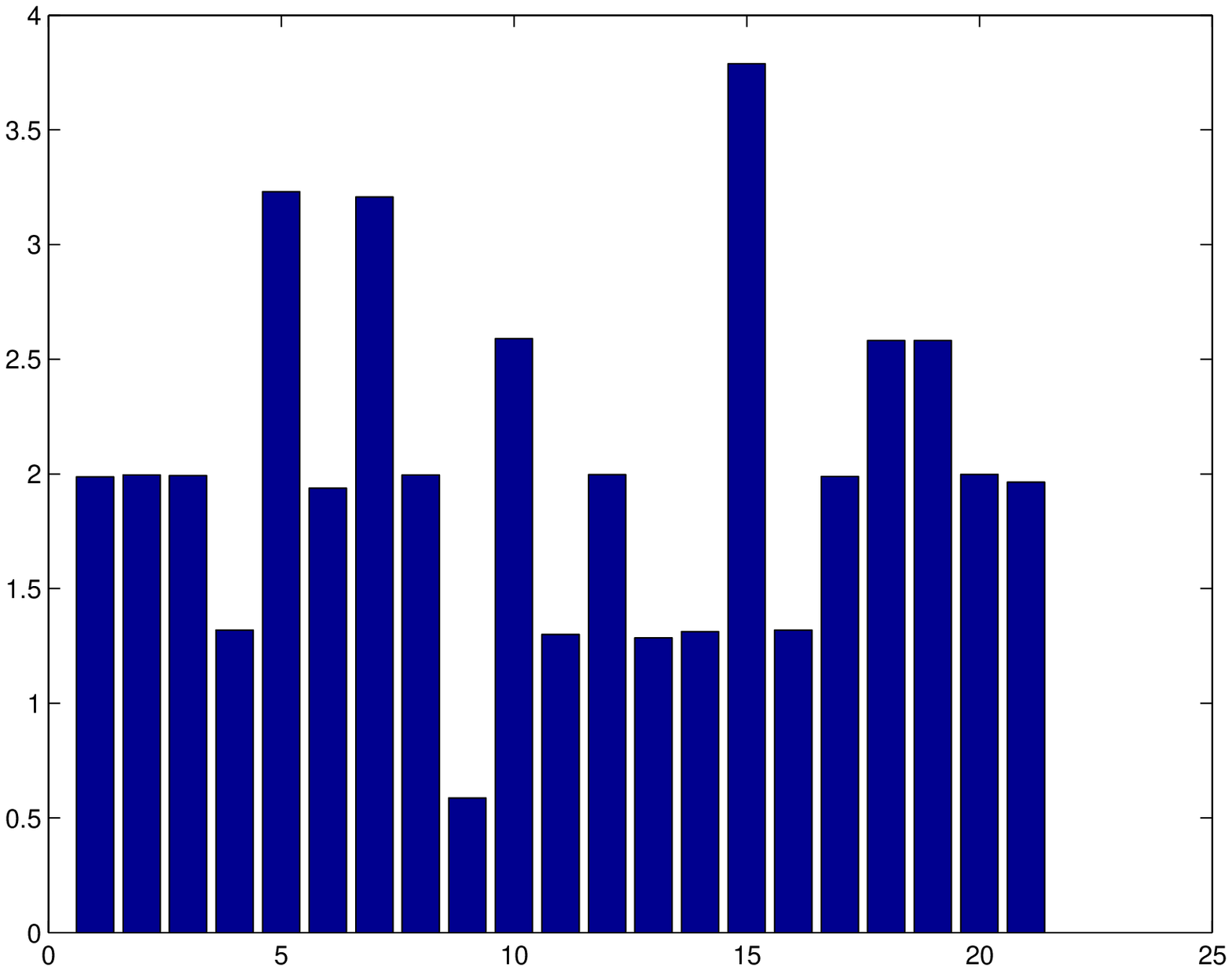}}
    \subfigure[q=0.4]{
    \label{Yeast-infect-local:a} 
    \centering
    \includegraphics[scale=0.25]{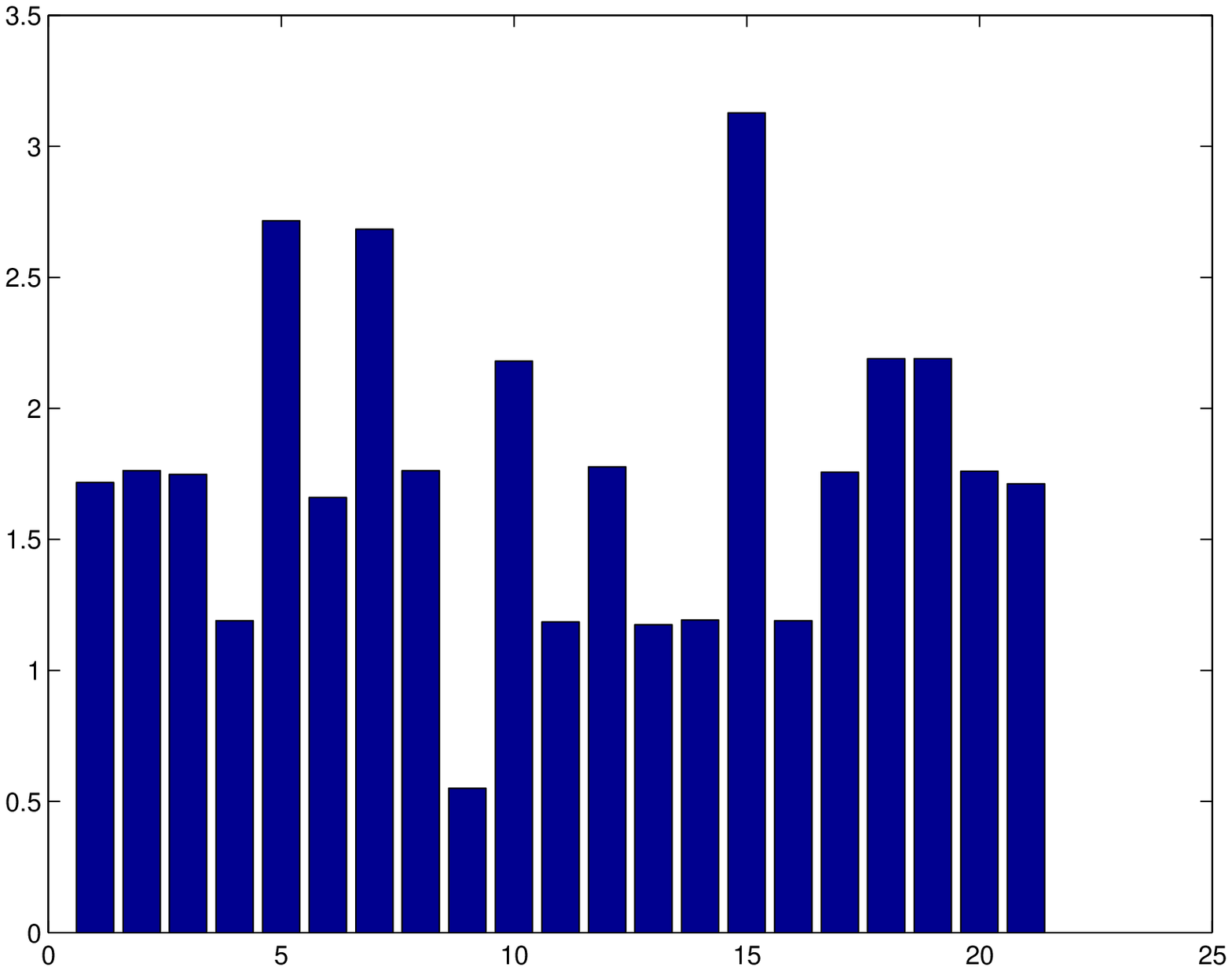}}
    \subfigure[q=0.5]{
    \label{Yeast-infect-local:b} 
    \centering
    \includegraphics[scale=0.25]{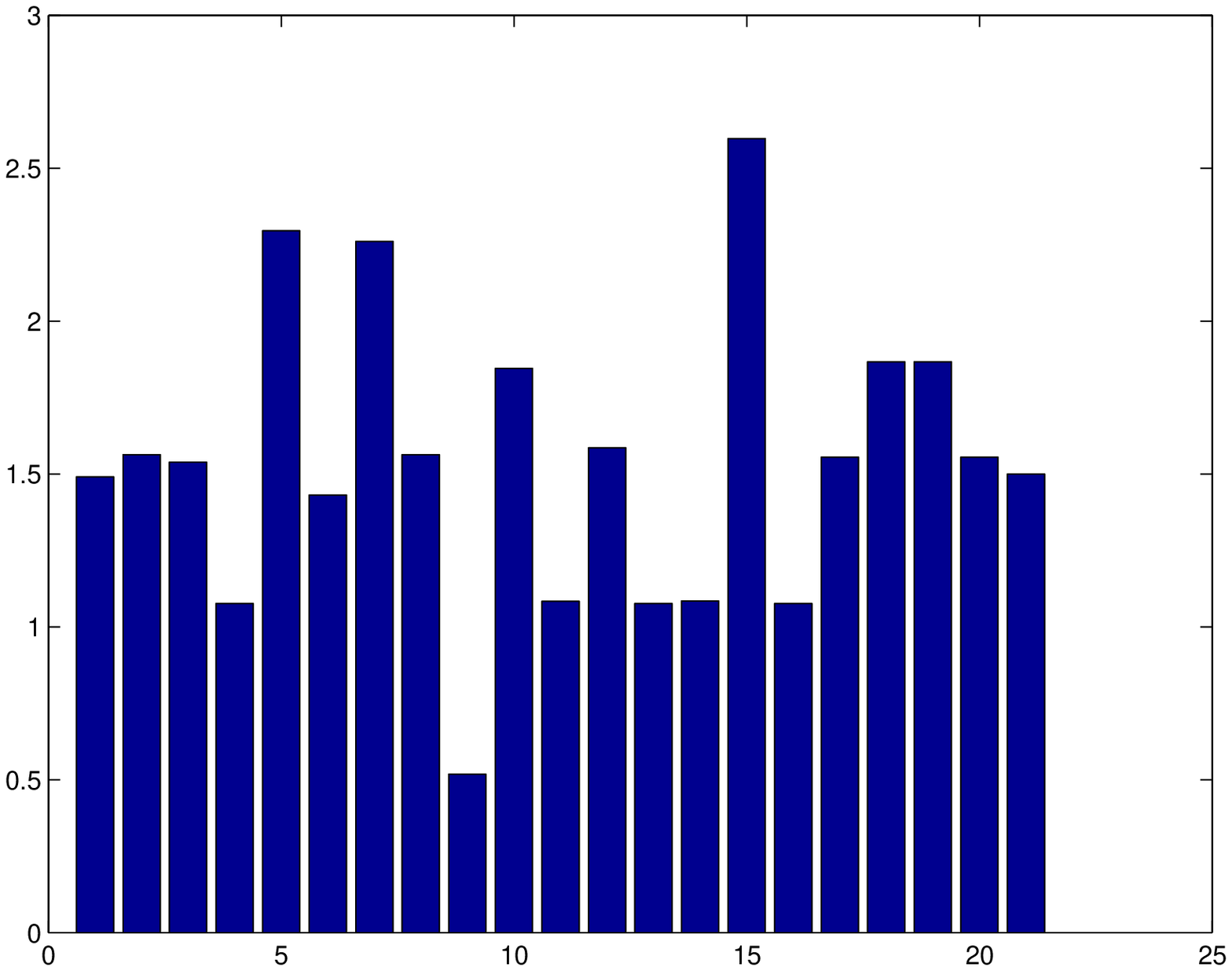}}
    \subfigure[q=0.6]{
    \label{Yeast-infect-local:b} 
    \centering
    \includegraphics[scale=0.25]{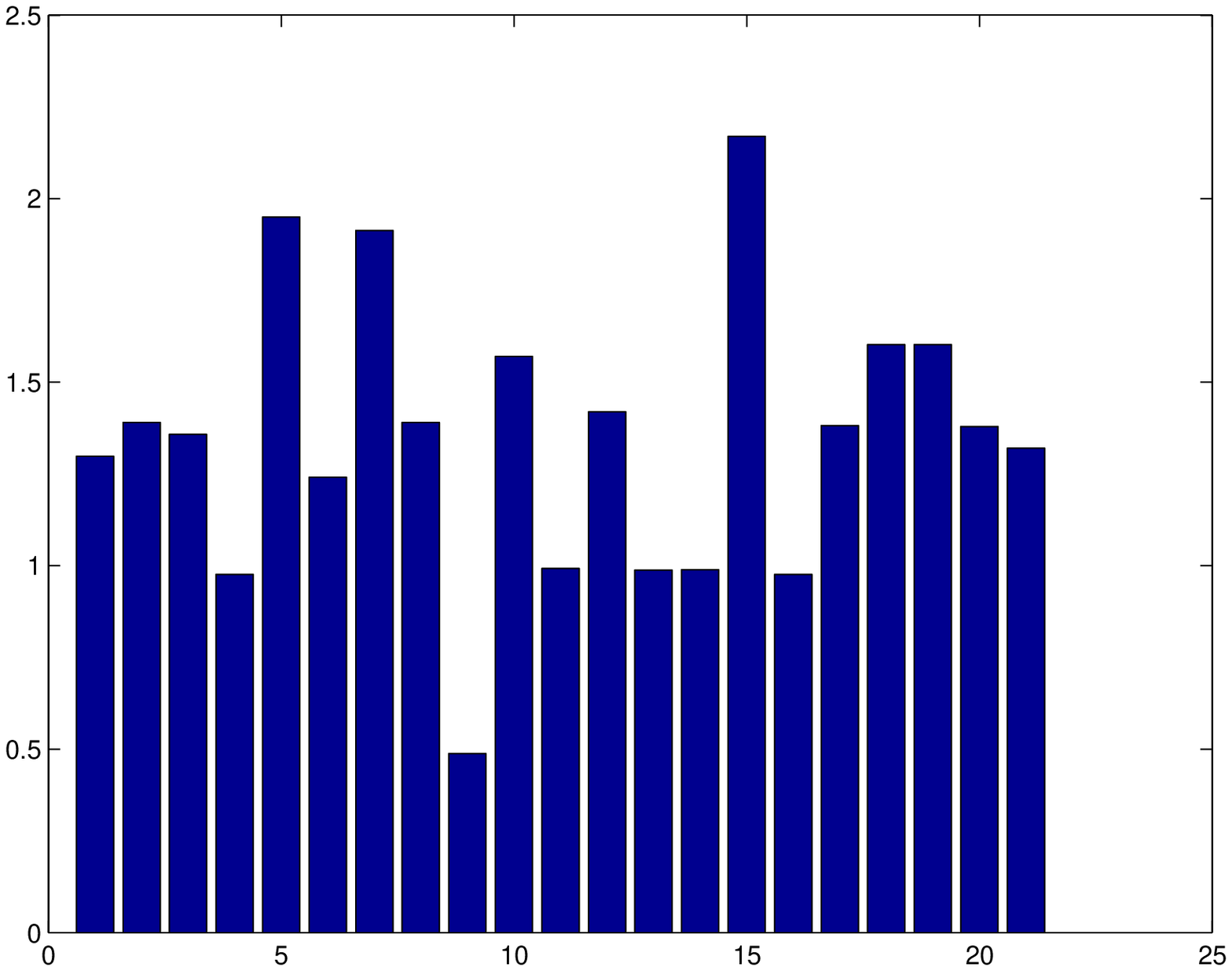}}
    \subfigure[q=0.7]{
    \label{Yeast-infect-local:b} 
    \centering
    \includegraphics[scale=0.25]{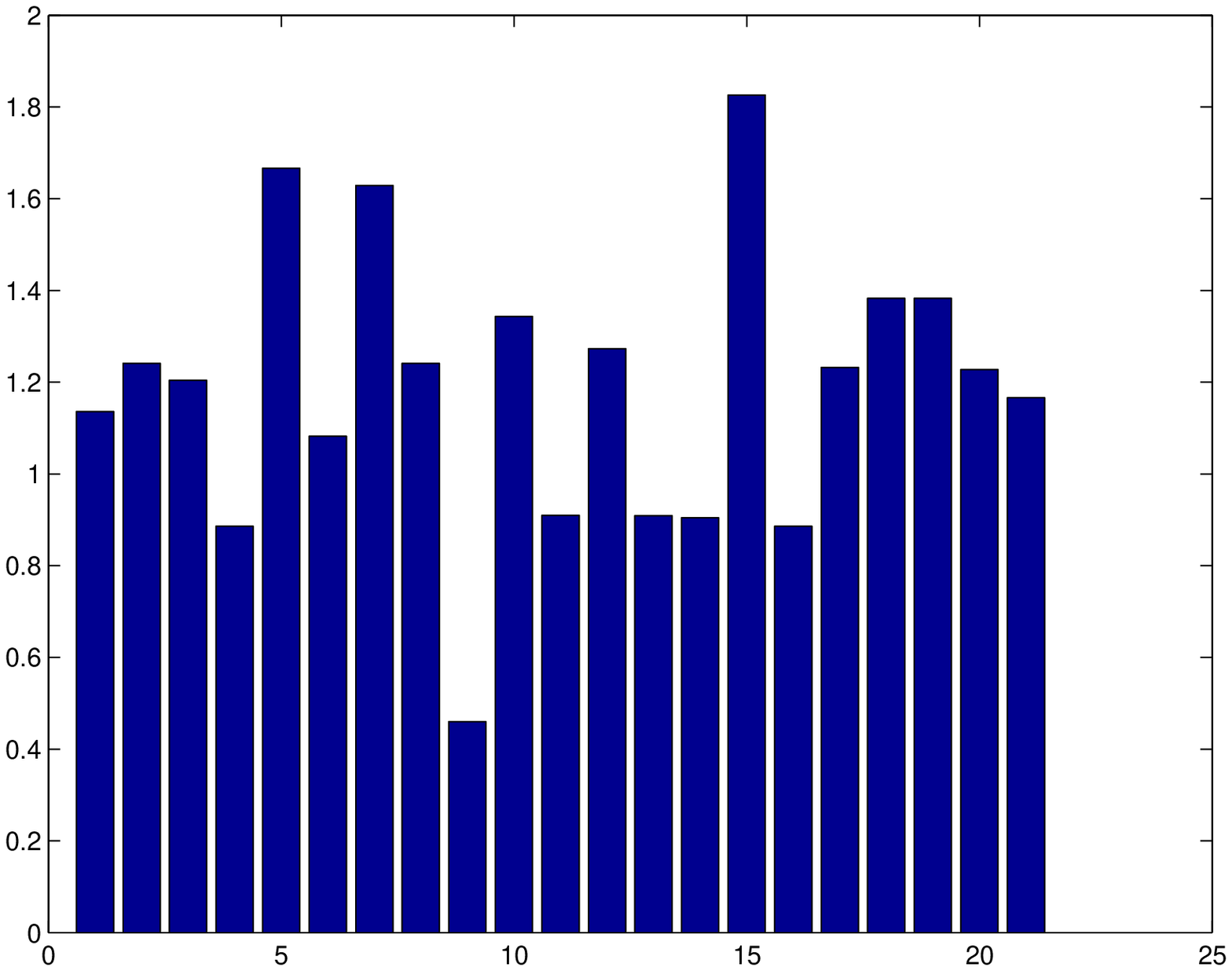}}
    \subfigure[q=0.8]{
    \label{Yeast-infect-local:b} 
    \centering
    \includegraphics[scale=0.25]{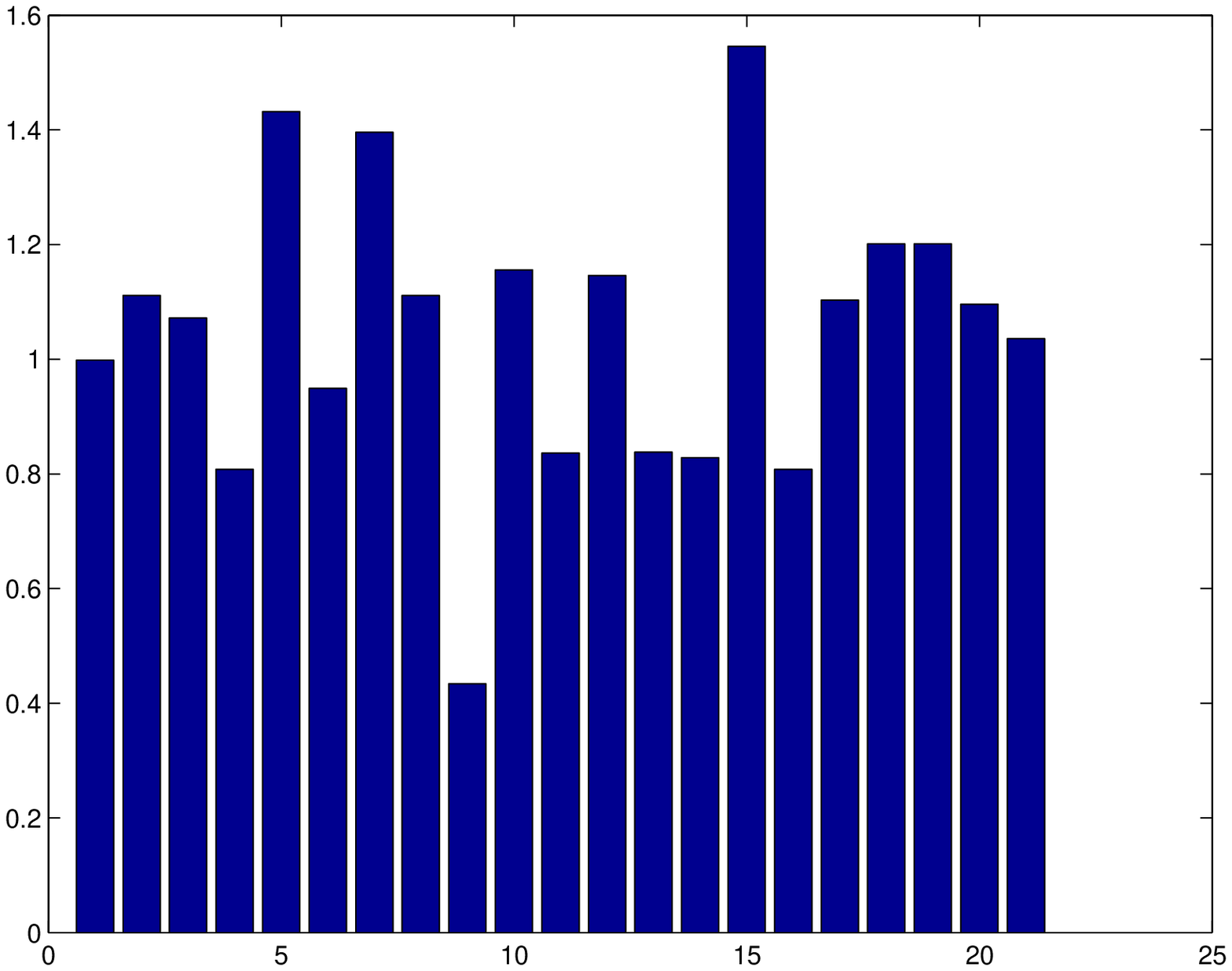}}
    \subfigure[q=0.9]{
    \label{Yeast-infect-local:b} 
    \centering
    \includegraphics[scale=0.25]{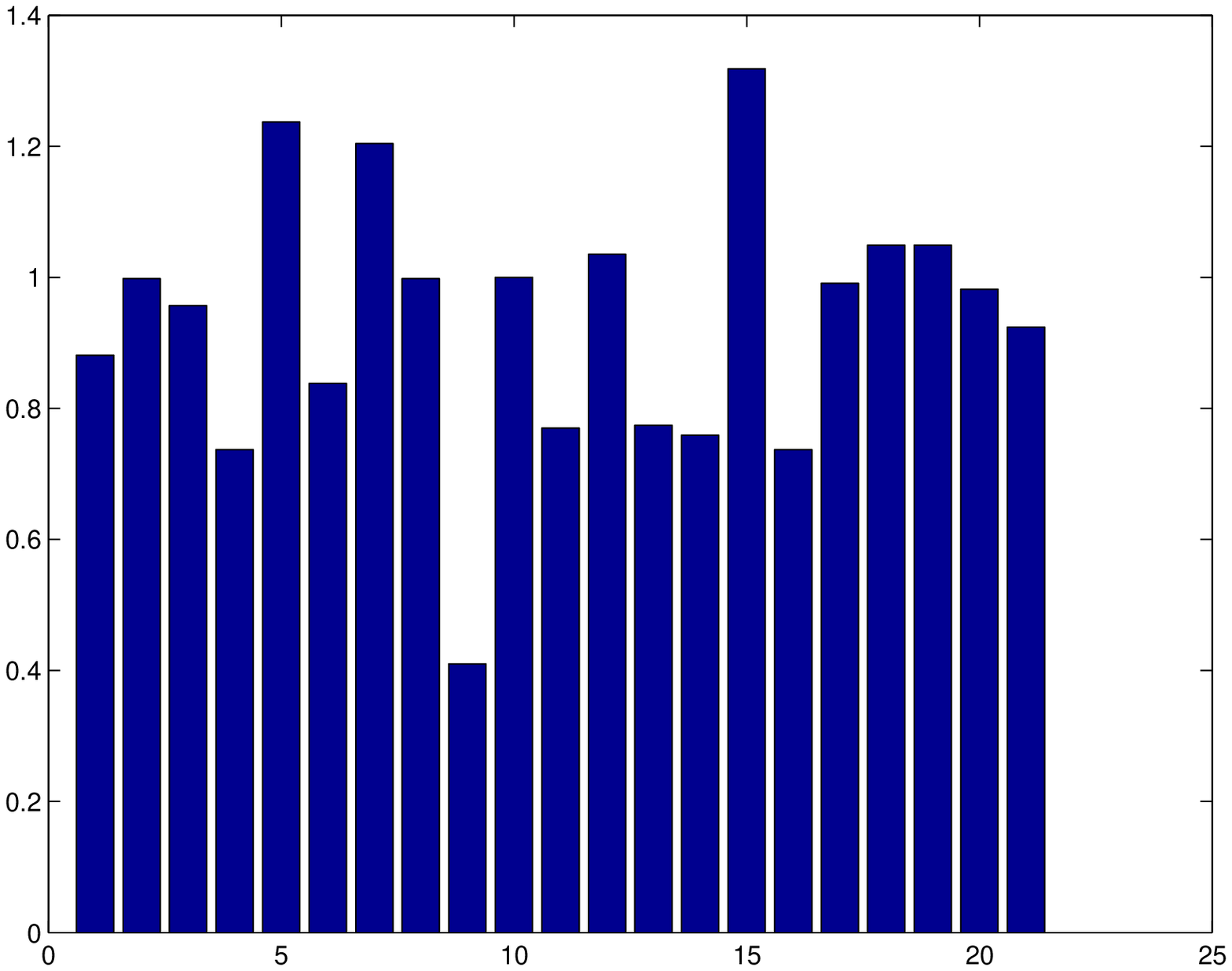}}
    \subfigure[q=1.0]{
    \label{Yeast-infect-local:b} 
    \centering
    \includegraphics[scale=0.25]{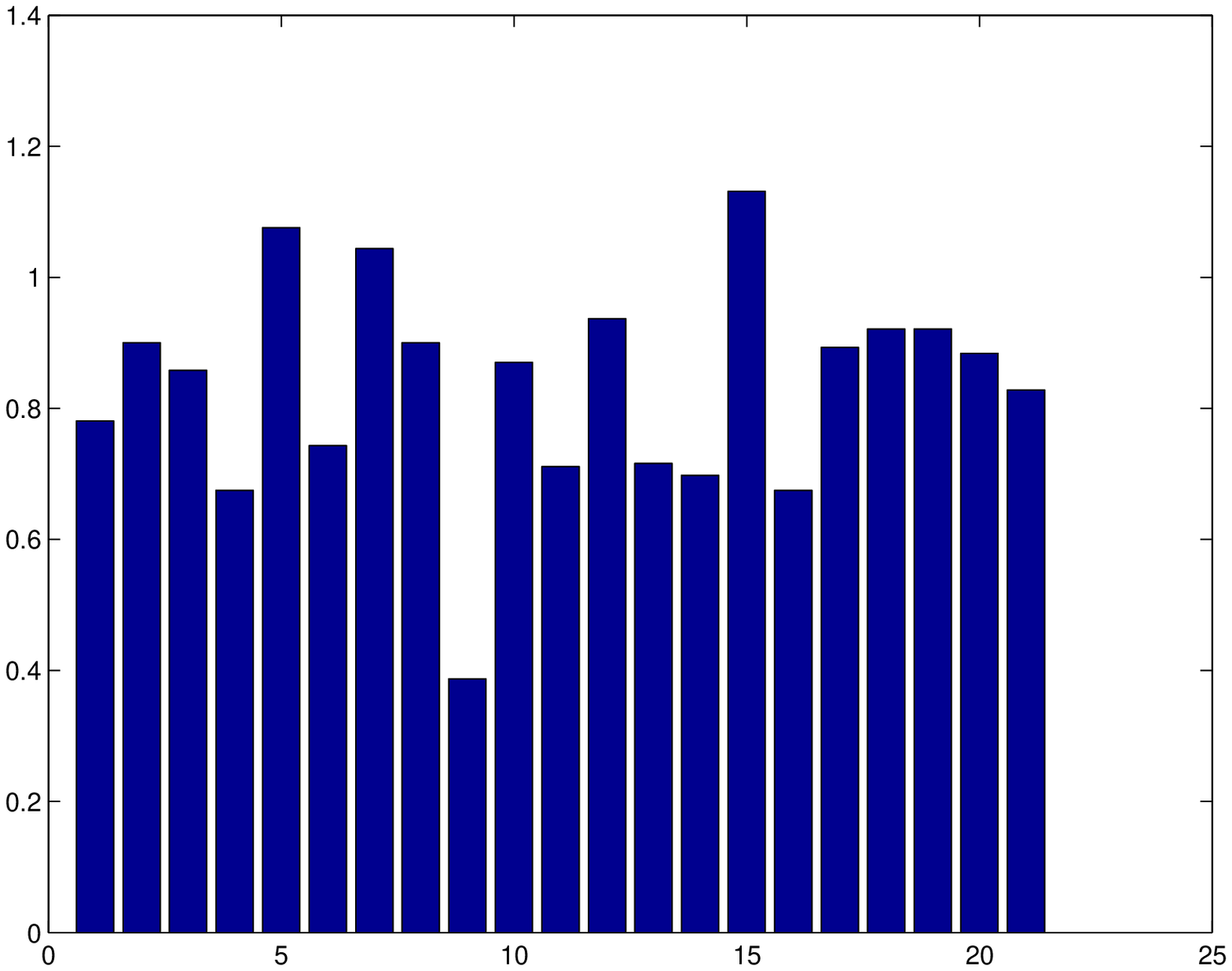}}
    \subfigure[q=1.1]{
    \label{Yeast-infect-local:b} 
    \centering
    \includegraphics[scale=0.25]{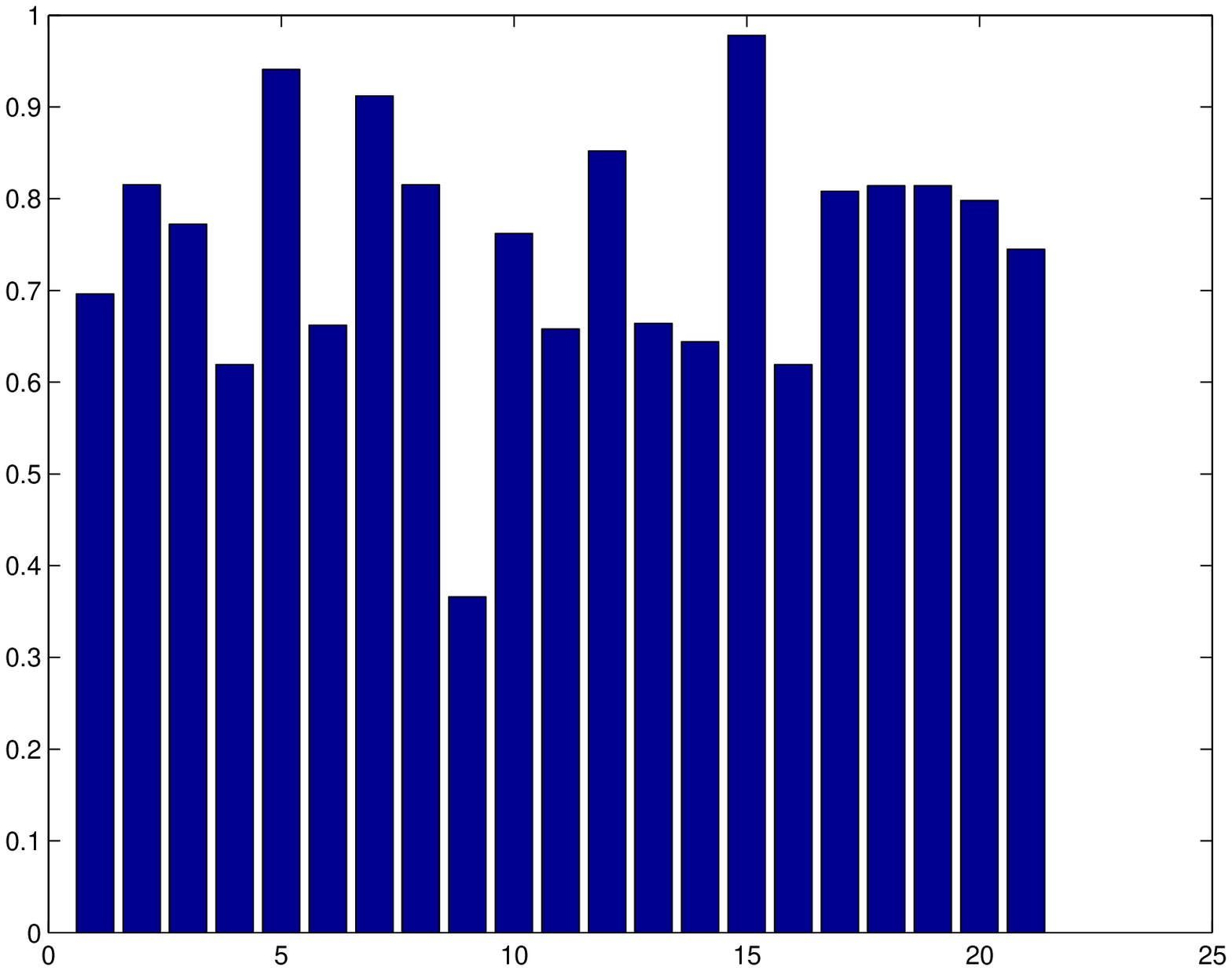}}
    \subfigure[q=1.2]{
    \label{Yeast-infect-local:b} 
    \centering
    \includegraphics[scale=0.25]{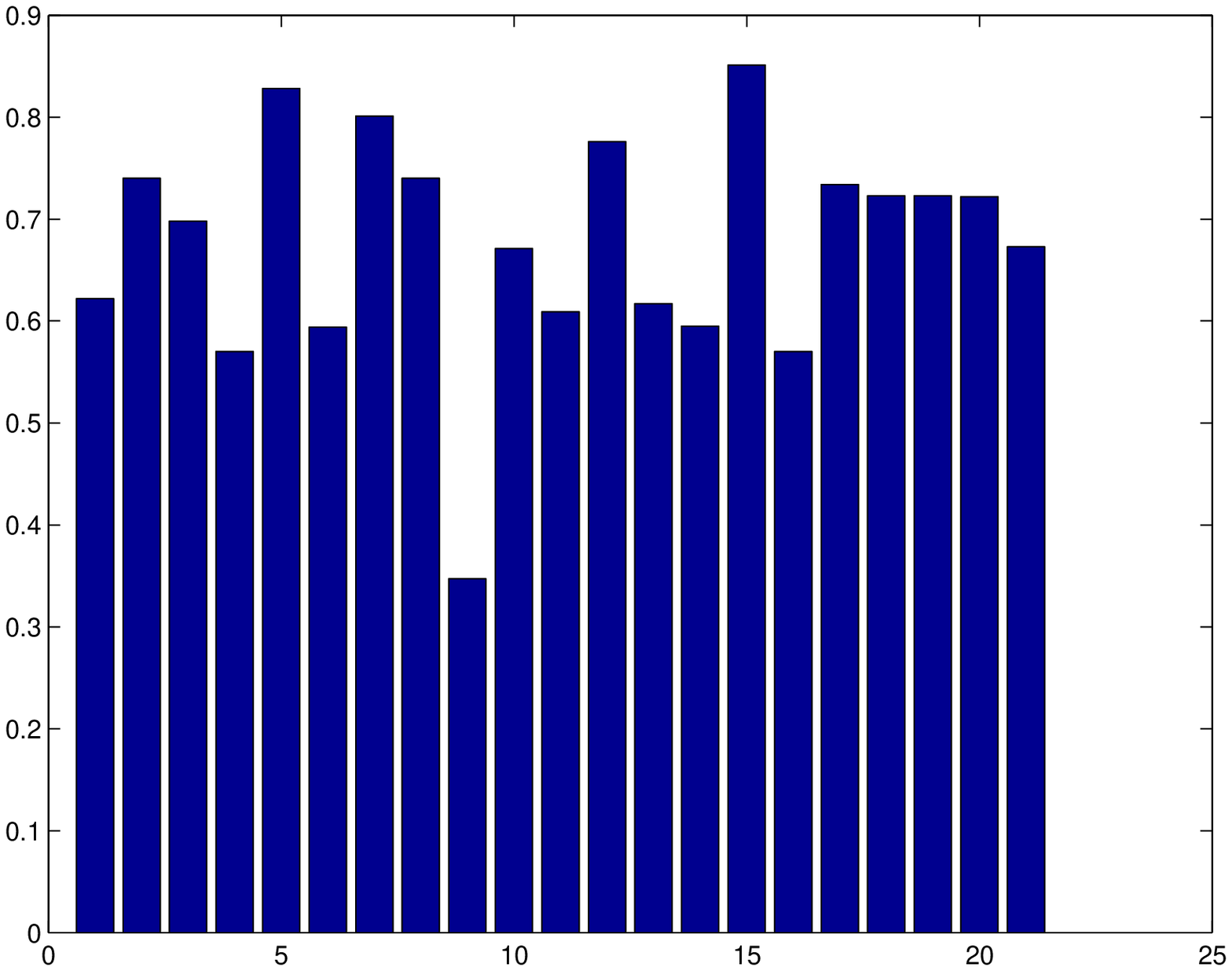}}
  \caption{The figure show the value of the nonextensive local structure entropy of each node in the example network A. The value of $q$ is big than 0.1 and small than 1.2. The caption of the subfigure show the value of $q$. The Abscissa in those subfigure represents the node's number and the ordinate represents the value of nonextensive local structure entropy. }\label{Yeast-infect-local}
\end{figure}

\begin{figure}
    \centering
    \subfigure[q=1.3]{
    \label{Yeast-infect-local:b} 
    \centering
    \includegraphics[scale=0.25]{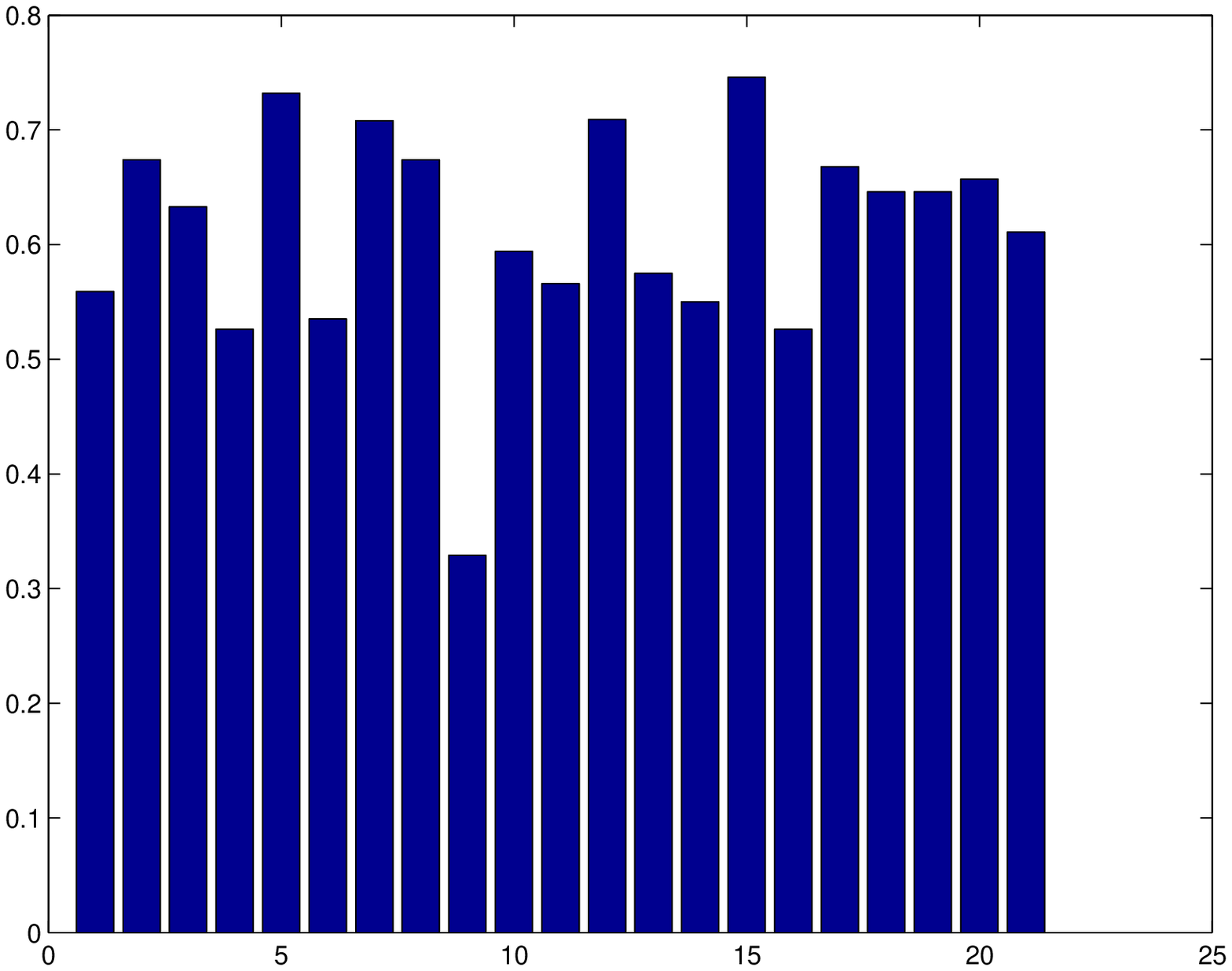}}
    \subfigure[q=1.4]{
    \label{Yeast-infect-local:b} 
    \centering
    \includegraphics[scale=0.25]{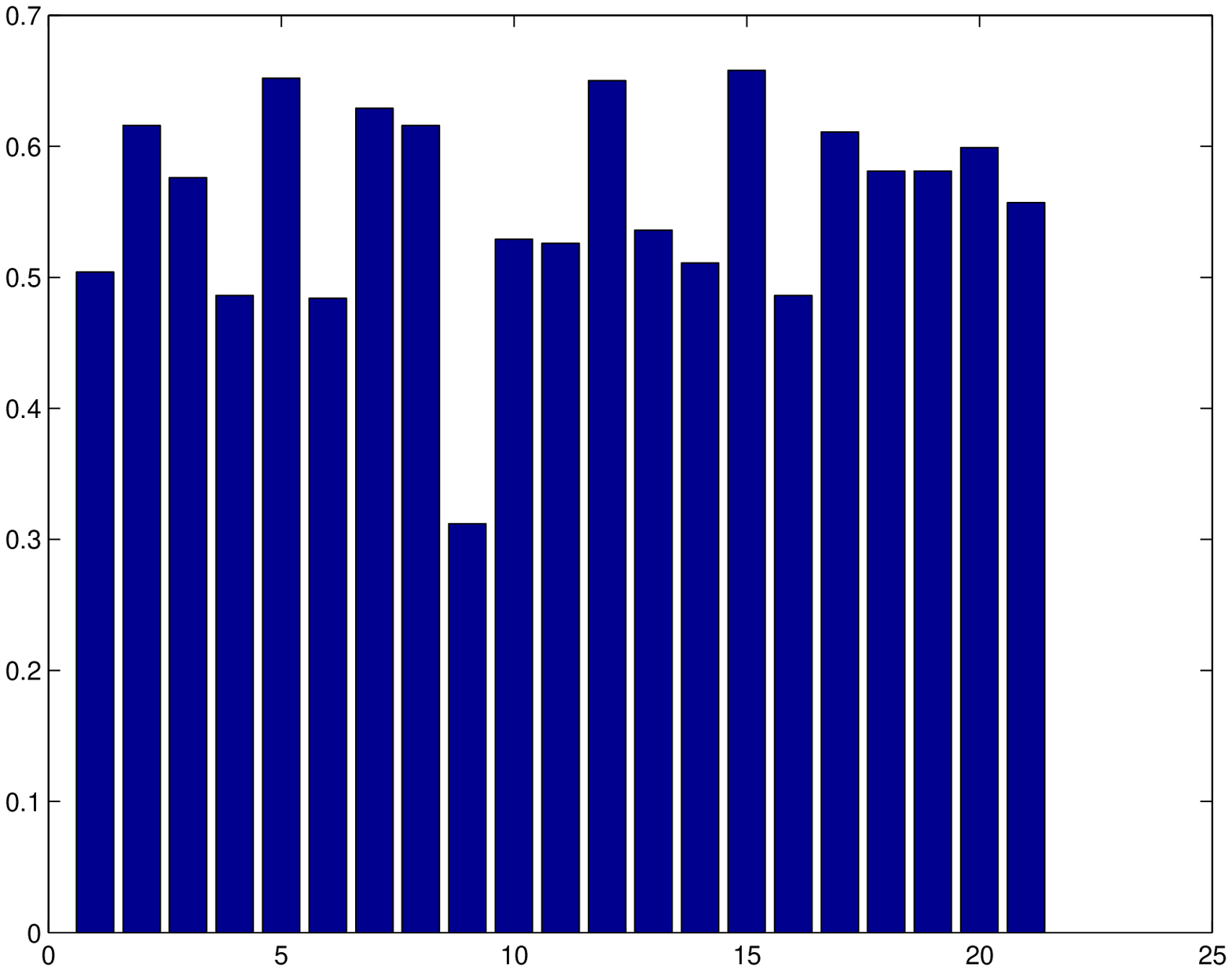}}
    \subfigure[q=1.5]{
    \label{Yeast-infect-local:b} 
    \centering
    \includegraphics[scale=0.25]{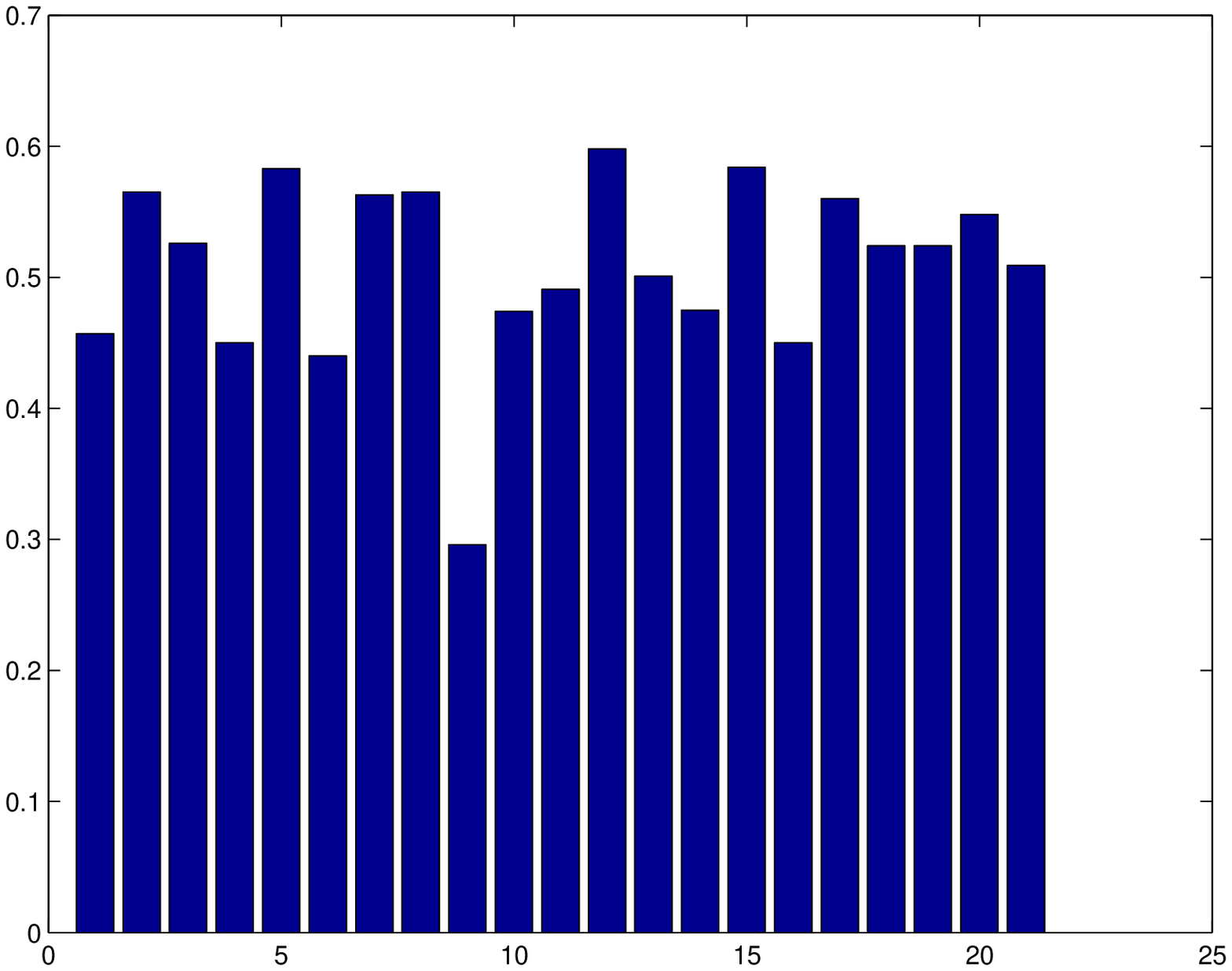}}
    \subfigure[q=1.6]{
    \label{Yeast-infect-local:b} 
    \centering
    \includegraphics[scale=0.25]{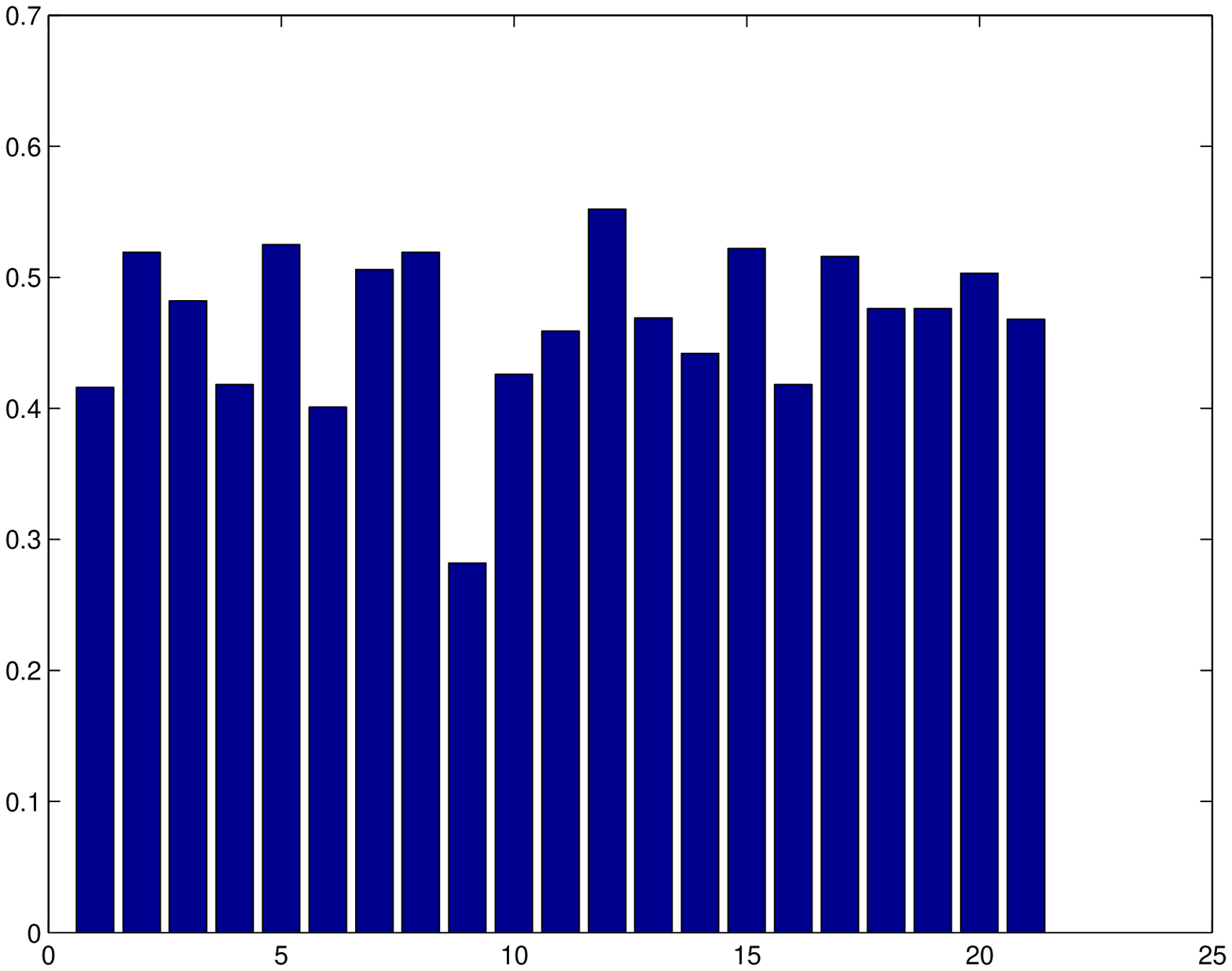}}
    \subfigure[q=1.7]{
    \label{Yeast-infect-local:b} 
    \centering
    \includegraphics[scale=0.25]{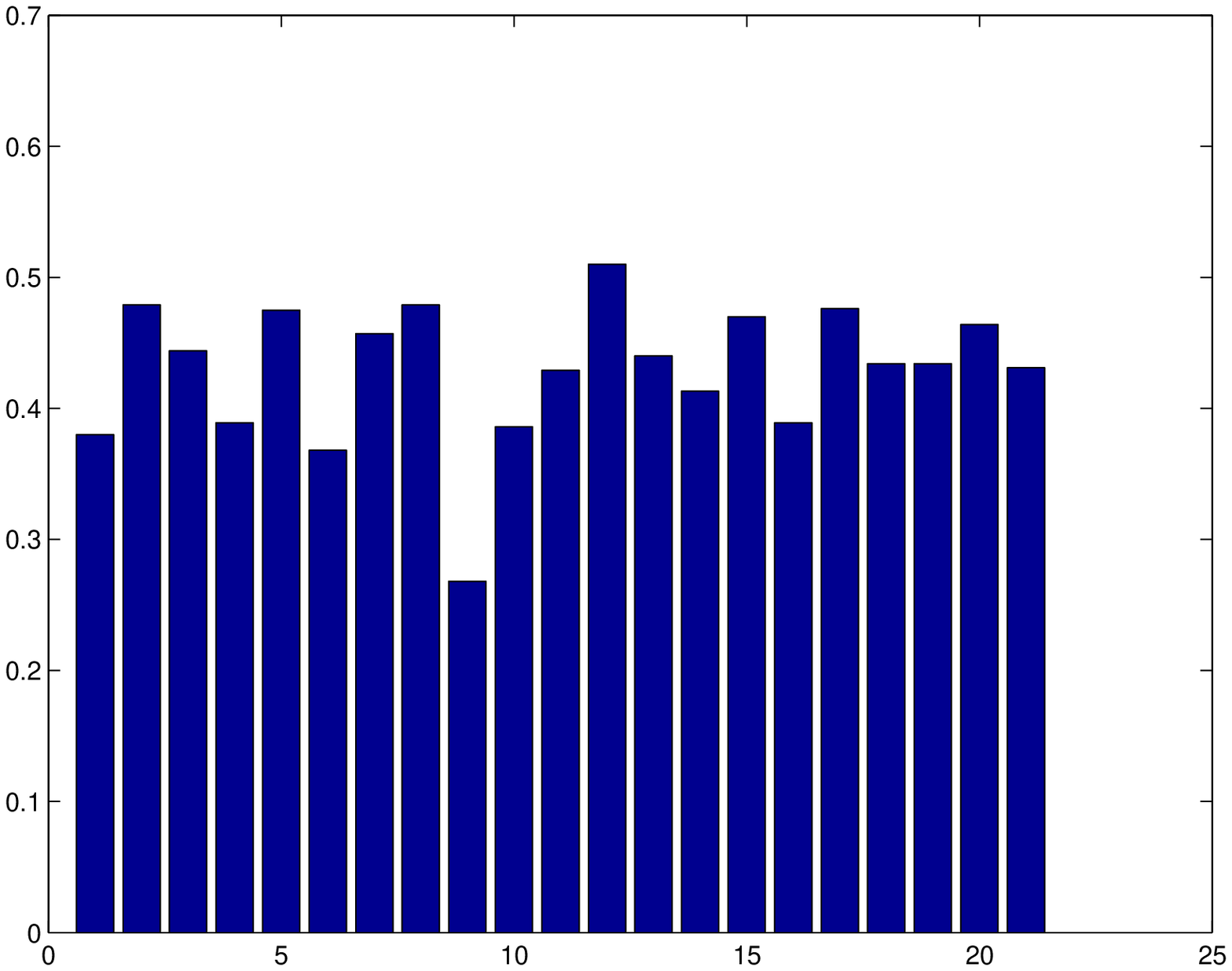}}
    \subfigure[q=1.8]{
    \label{Yeast-infect-local:b} 
    \centering
    \includegraphics[scale=0.25]{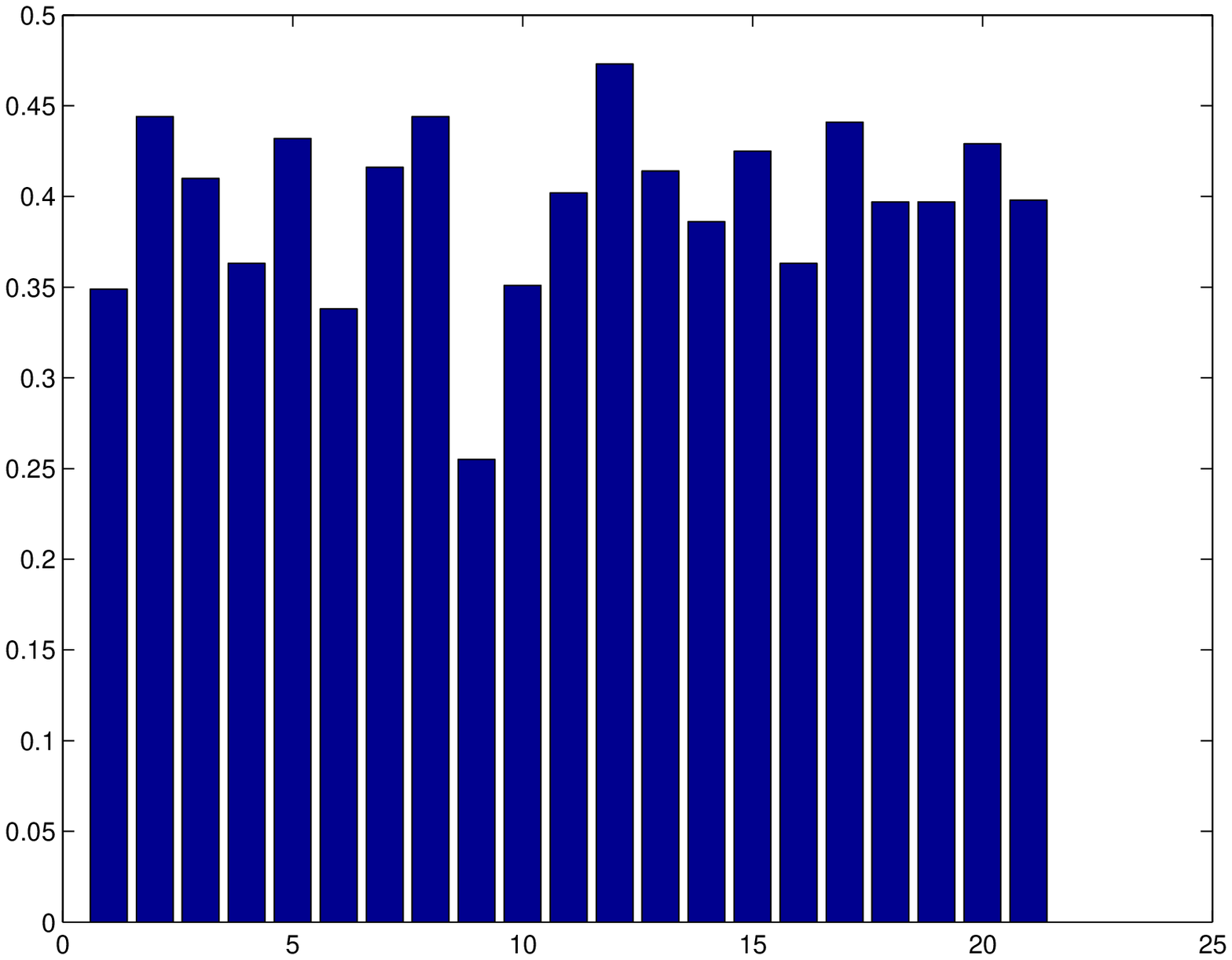}}
    \subfigure[q=1.9]{
    \label{Yeast-infect-local:b} 
    \centering
    \includegraphics[scale=0.25]{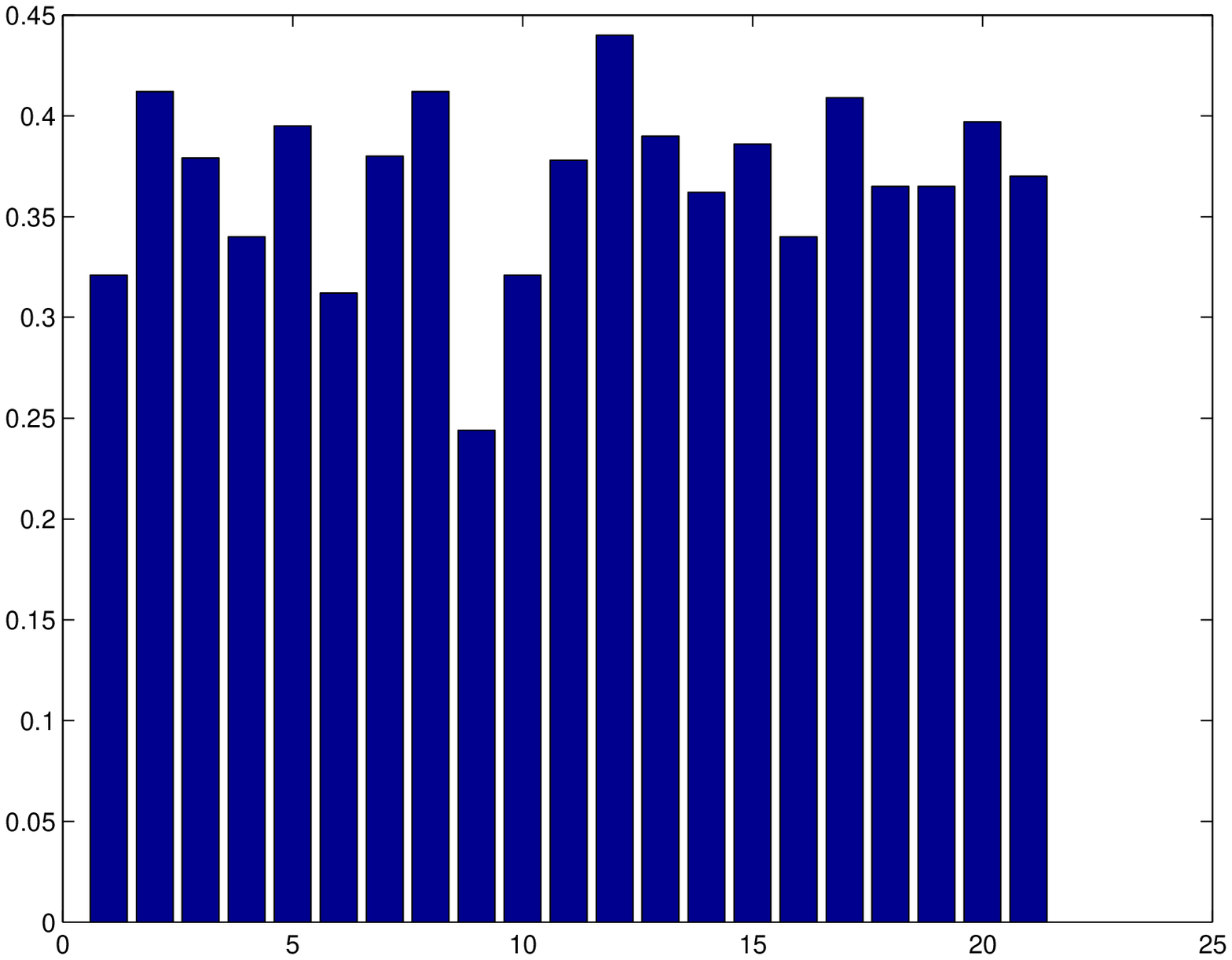}}
    \subfigure[q=2.0]{
    \label{Yeast-infect-local:b} 
    \centering
    \includegraphics[scale=0.25]{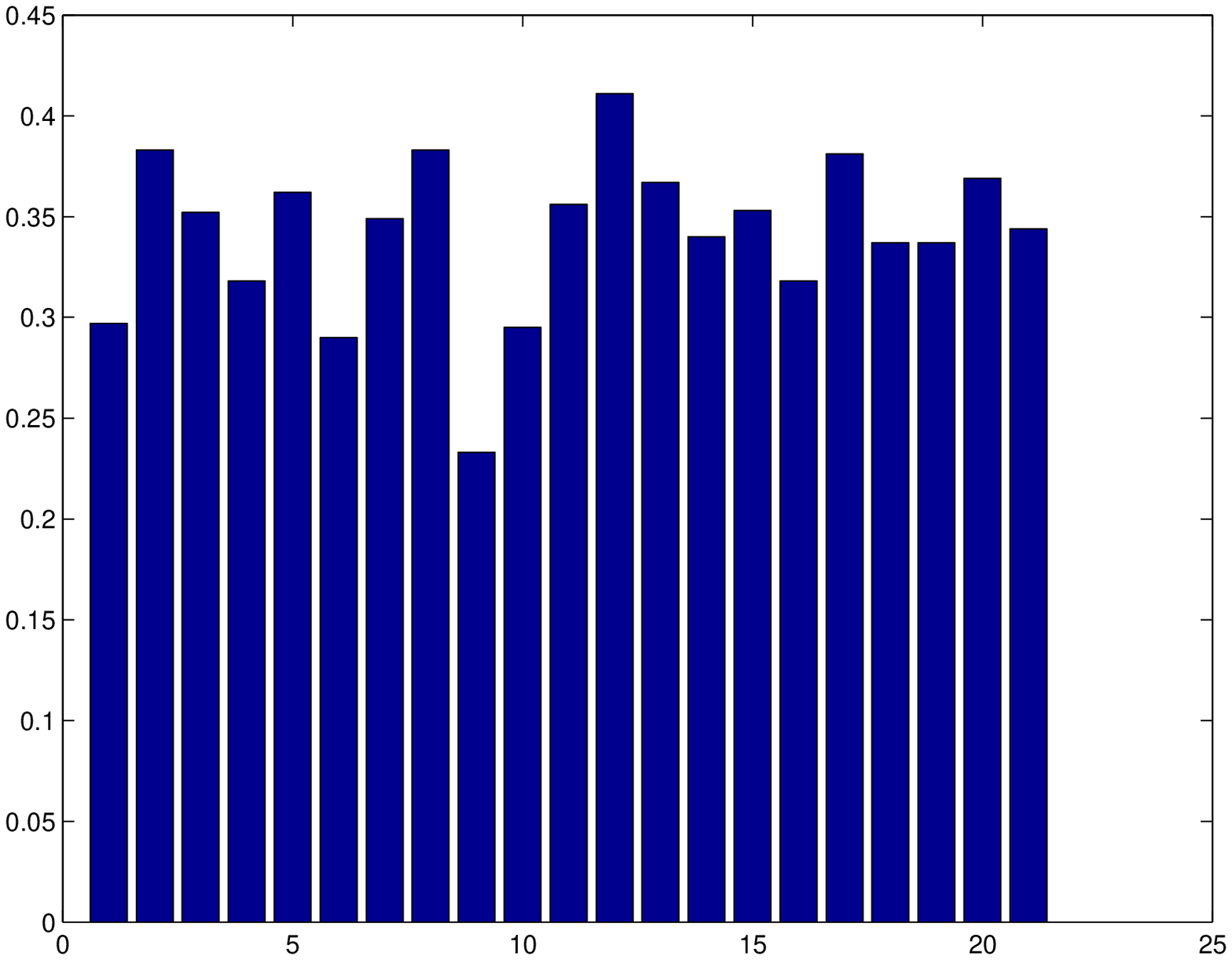}}
    \subfigure[q=2.2]{
    \label{Yeast-infect-local:b} 
    \centering
    \includegraphics[scale=0.25]{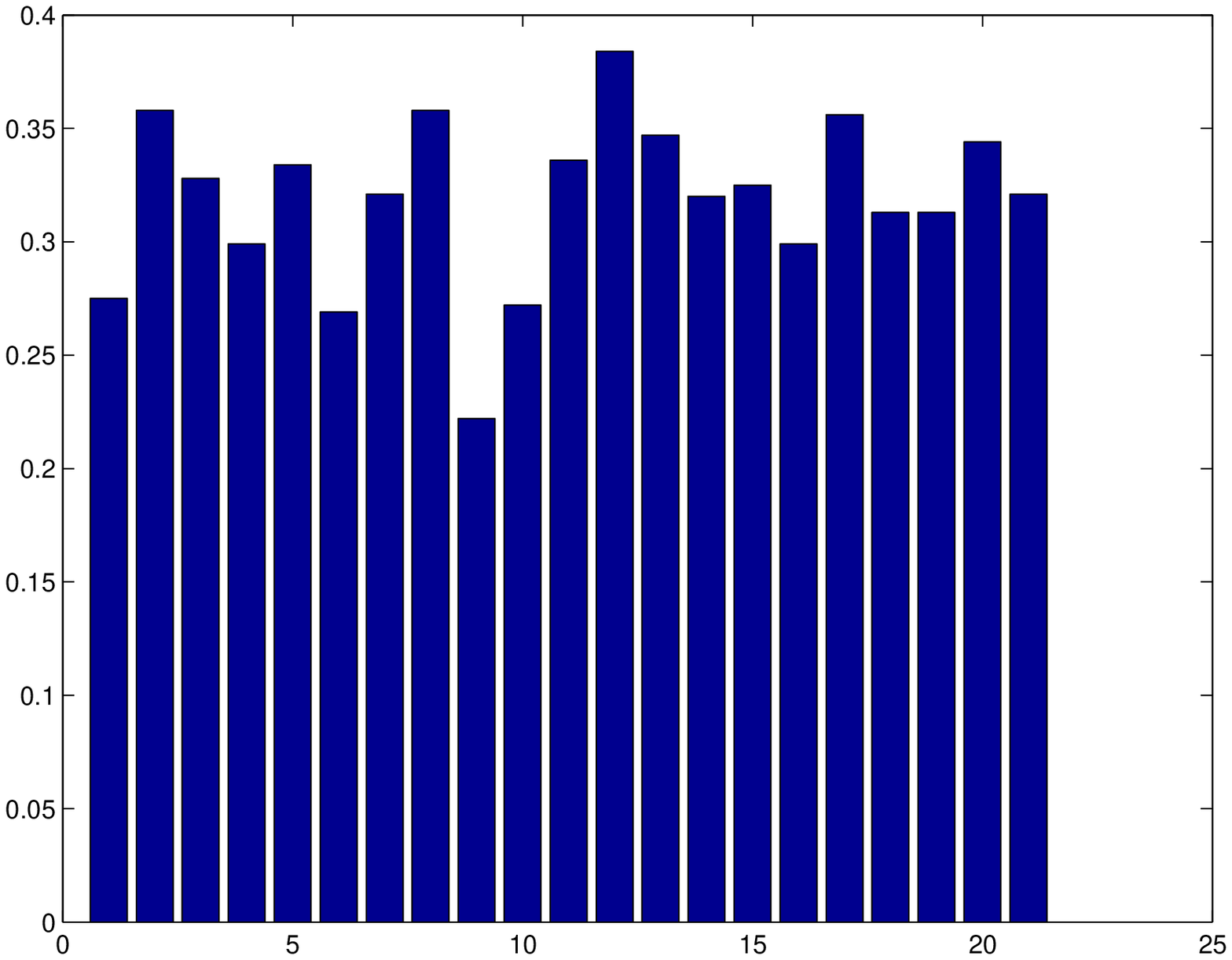}}
    \subfigure[q=2.4]{
    \label{Yeast-infect-local:b} 
    \centering
    \includegraphics[scale=0.25]{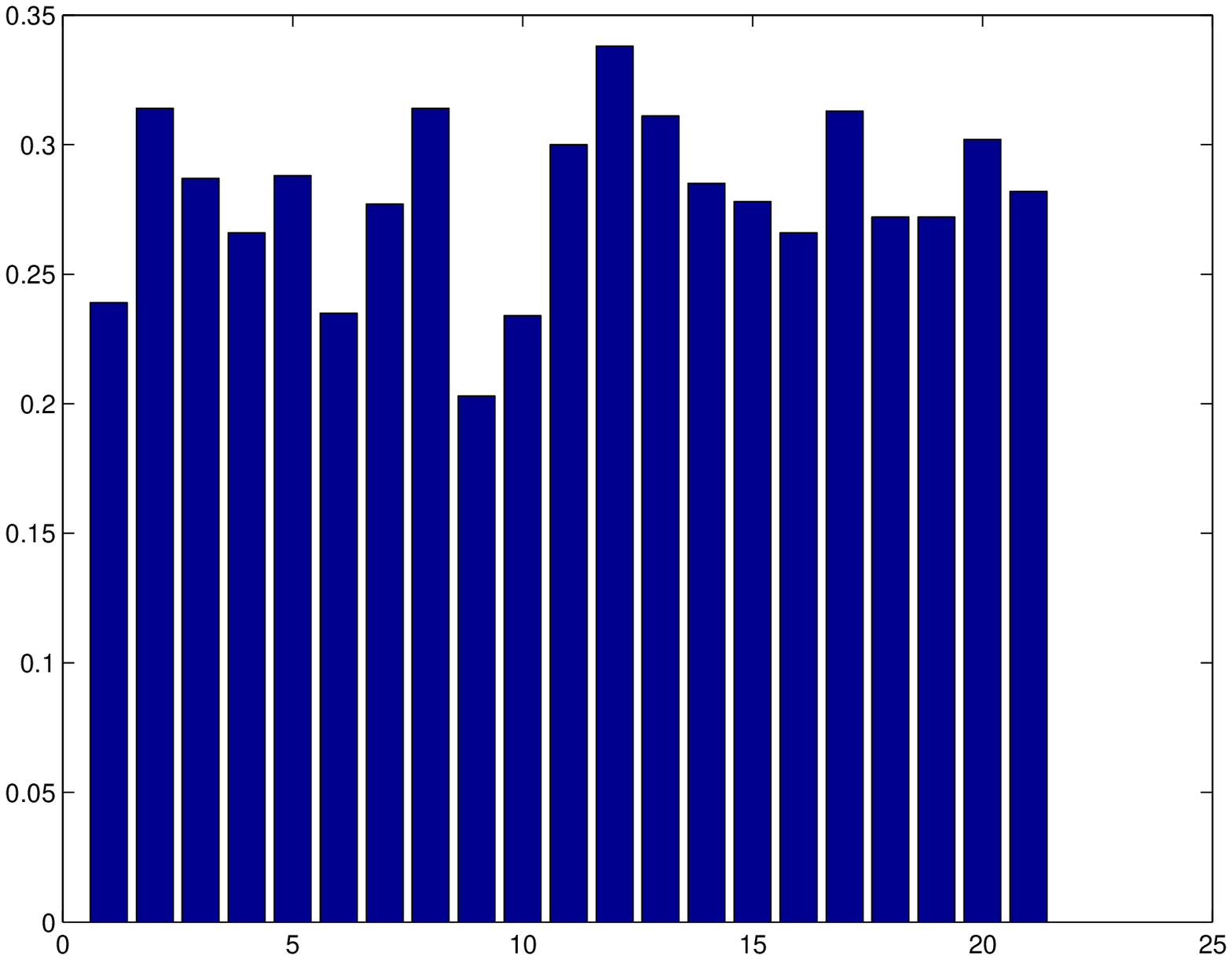}}
    \subfigure[q=2.6]{
    \label{Yeast-infect-local:b} 
    \centering
    \includegraphics[scale=0.25]{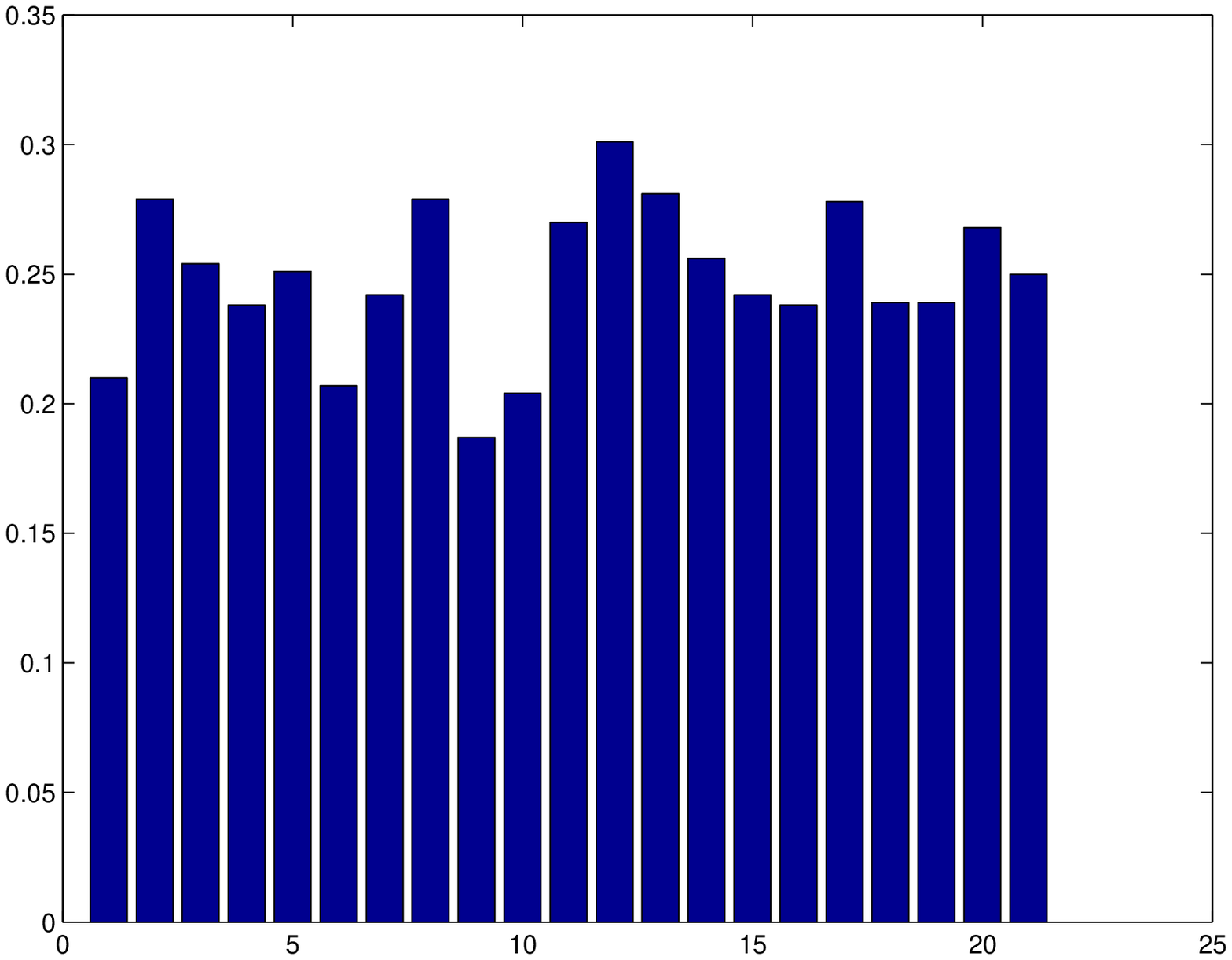}}
    \subfigure[q=2.8]{
    \label{Yeast-infect-local:b} 
    \centering
    \includegraphics[scale=0.25]{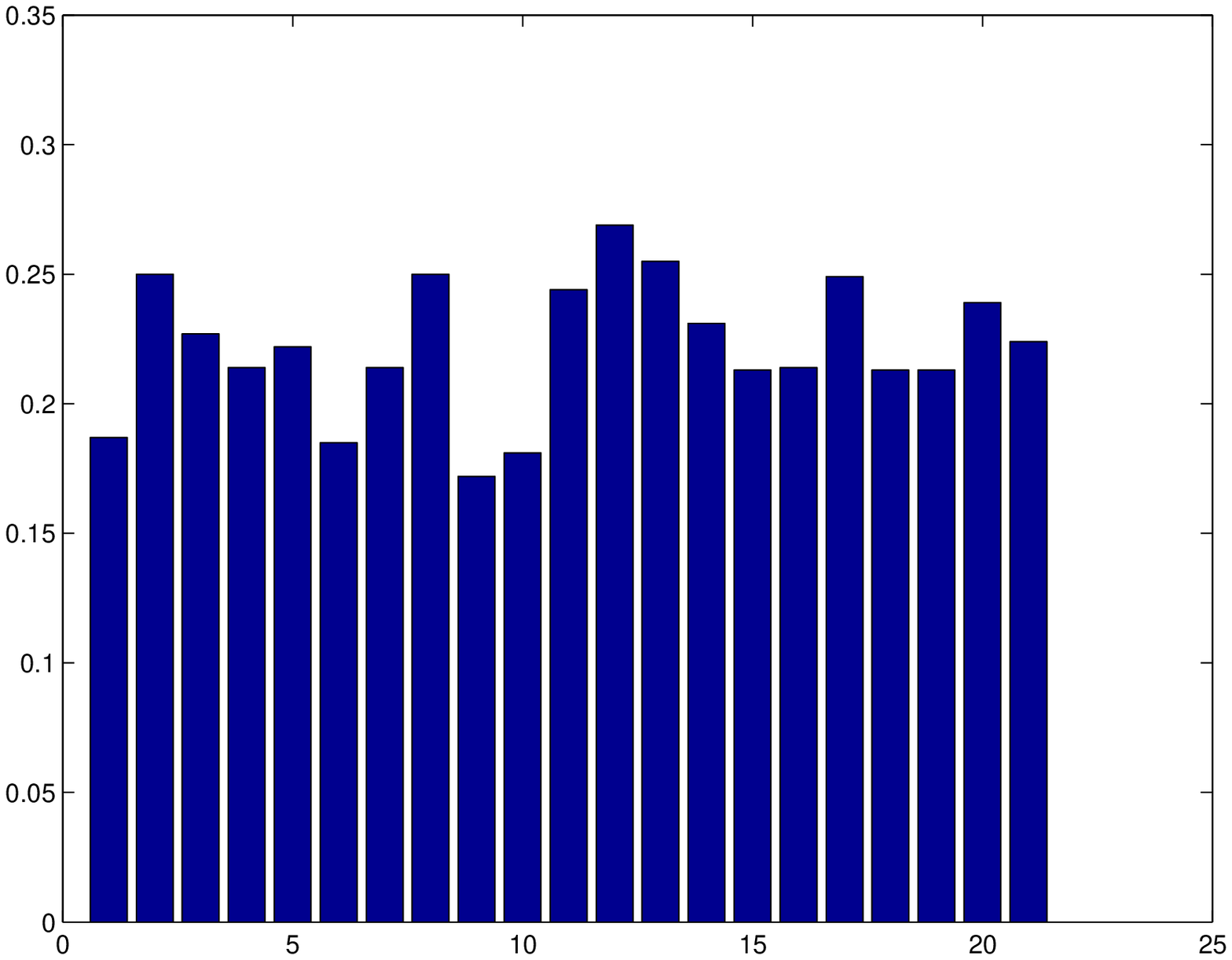}}
  \caption{The figure show the value of the nonextensive local structure entropy of each node in the example network A. The value of $q$ is big than 1.3 and small than 2.8. The caption of the subfigure show the value of $q$. The Abscissa in those subfigure represents the node's number and the ordinate represents the value of nonextensive local structure entropy. }\label{Yeast-infect-local}
\end{figure}

\begin{figure}
    \centering
    \subfigure[q=3.0]{
    \label{Yeast-infect-local:b} 
    \centering
    \includegraphics[scale=0.25]{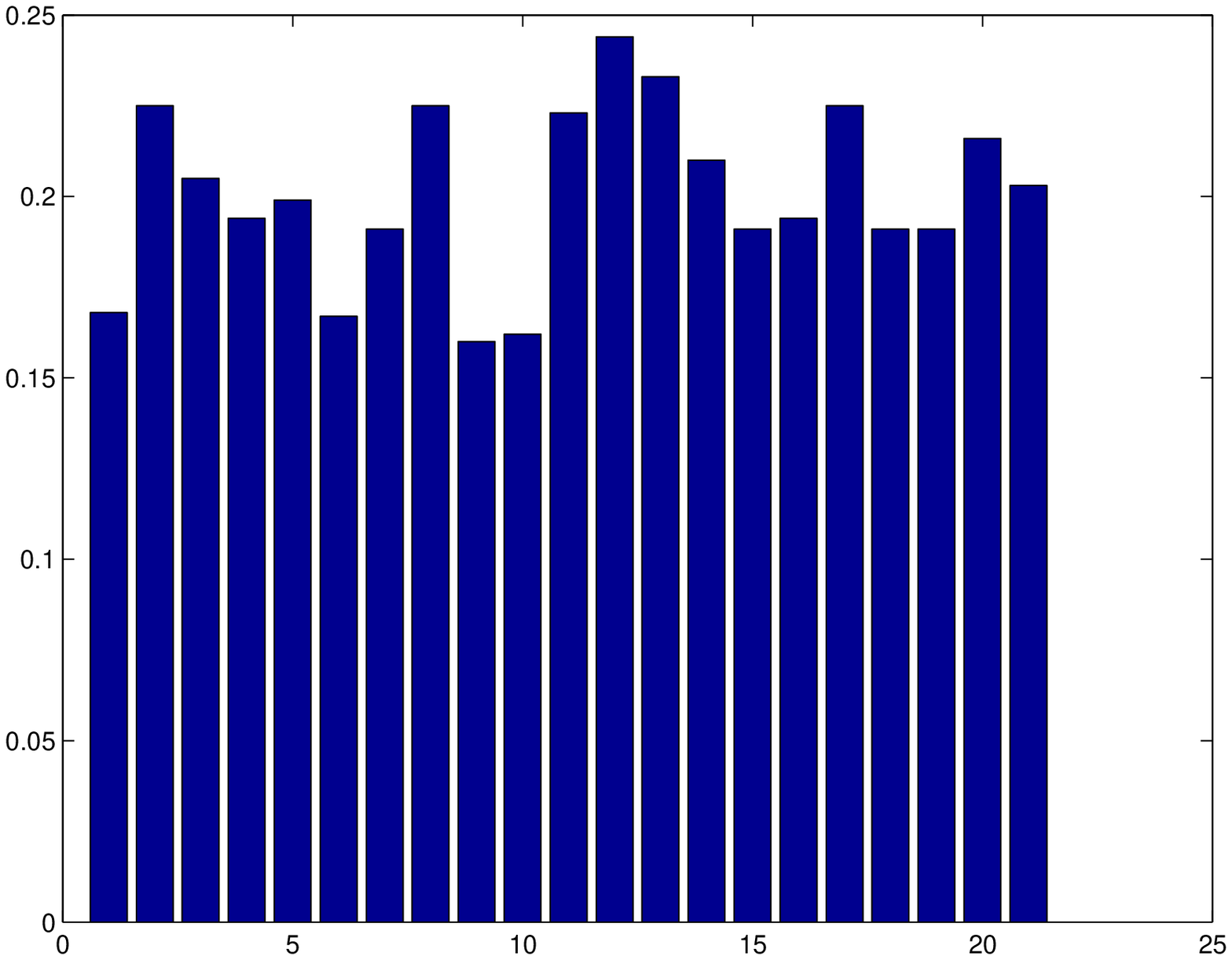}}
    \subfigure[q=3.2]{
    \label{Yeast-infect-local:b} 
    \centering
    \includegraphics[scale=0.25]{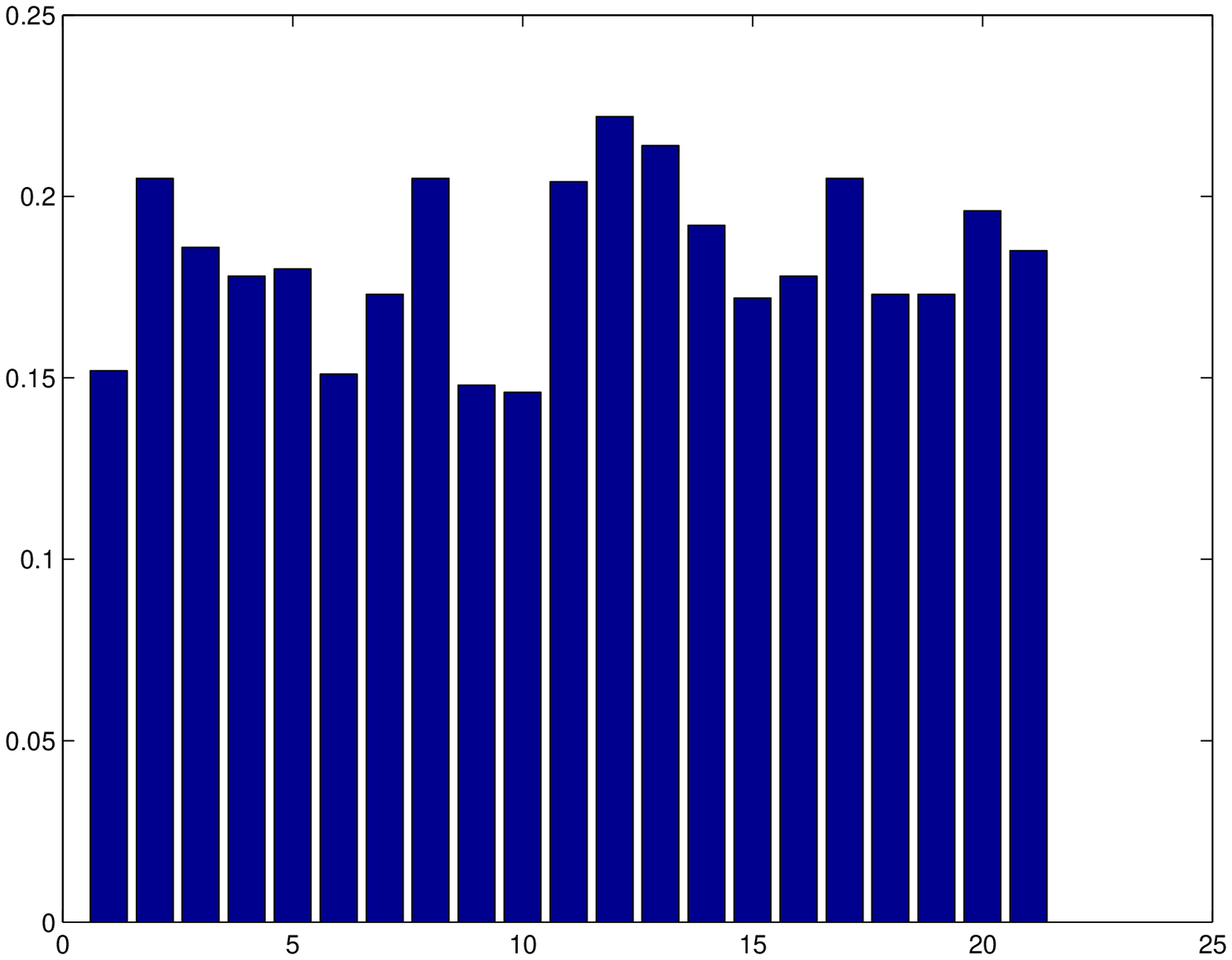}}
    \subfigure[q=3.4]{
    \label{Yeast-infect-local:b} 
    \centering
    \includegraphics[scale=0.25]{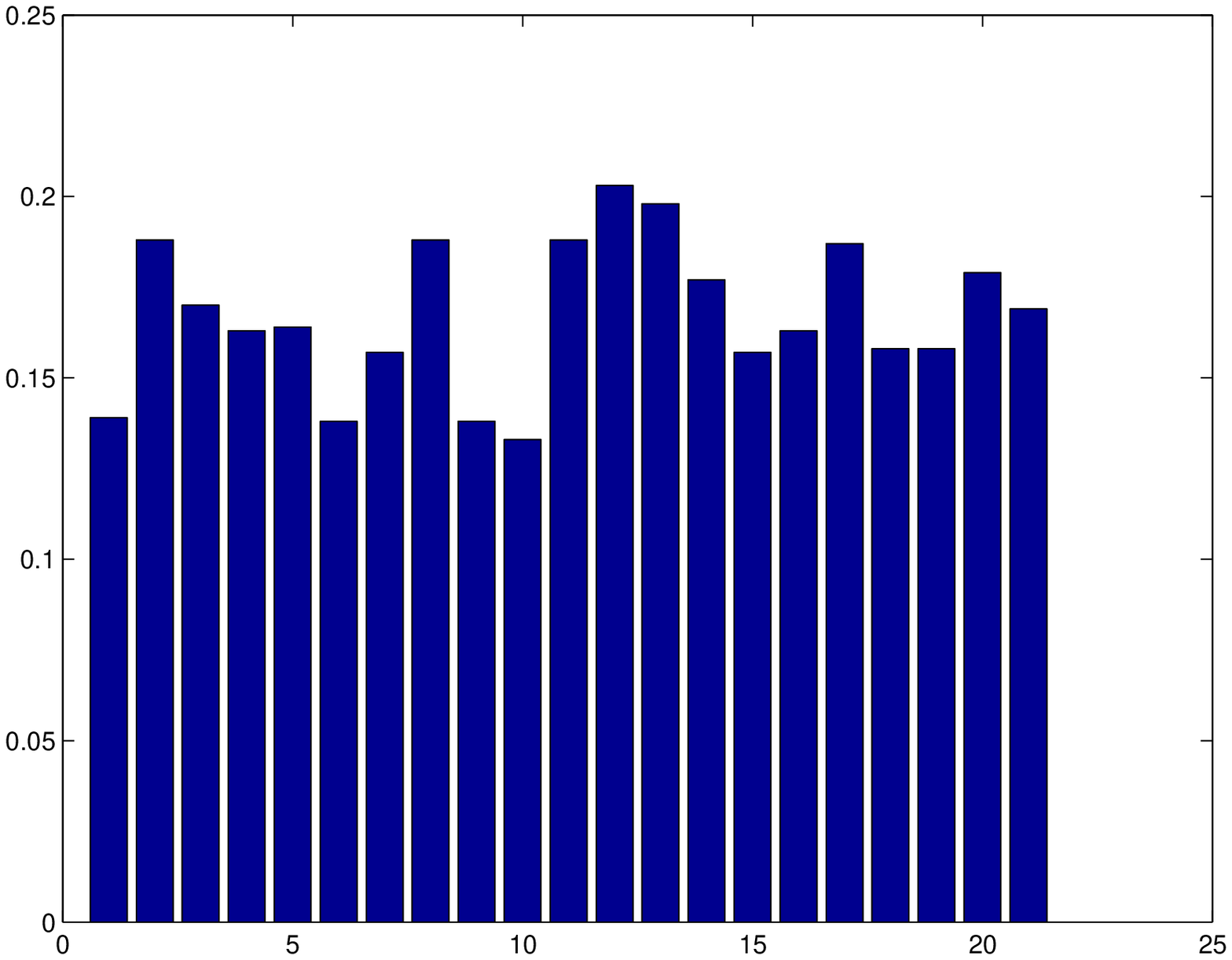}}
    \subfigure[q=3.6]{
    \label{Yeast-infect-local:b} 
    \centering
    \includegraphics[scale=0.25]{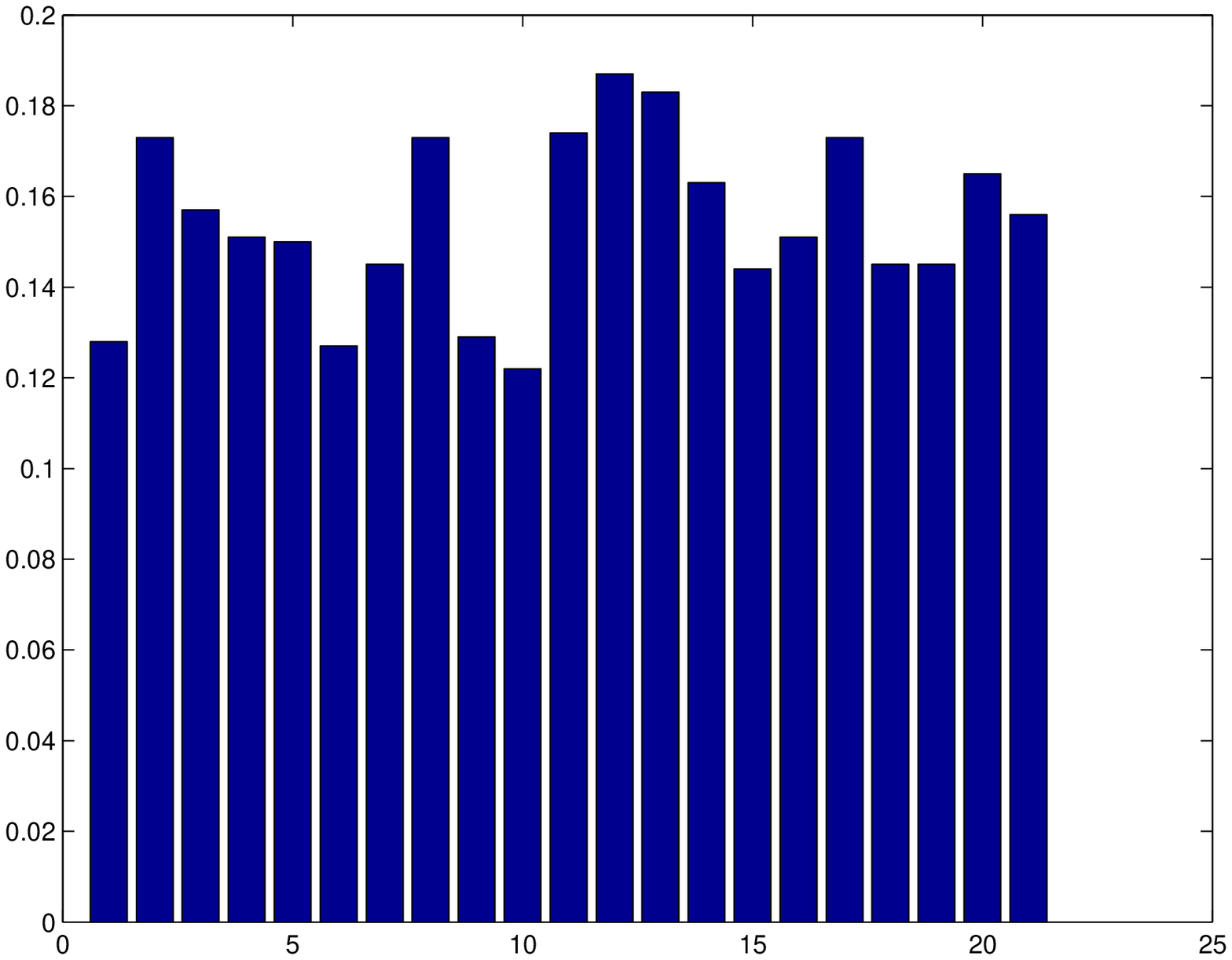}}
    \subfigure[q=3.8]{
    \label{Yeast-infect-local:b} 
    \centering
    \includegraphics[scale=0.25]{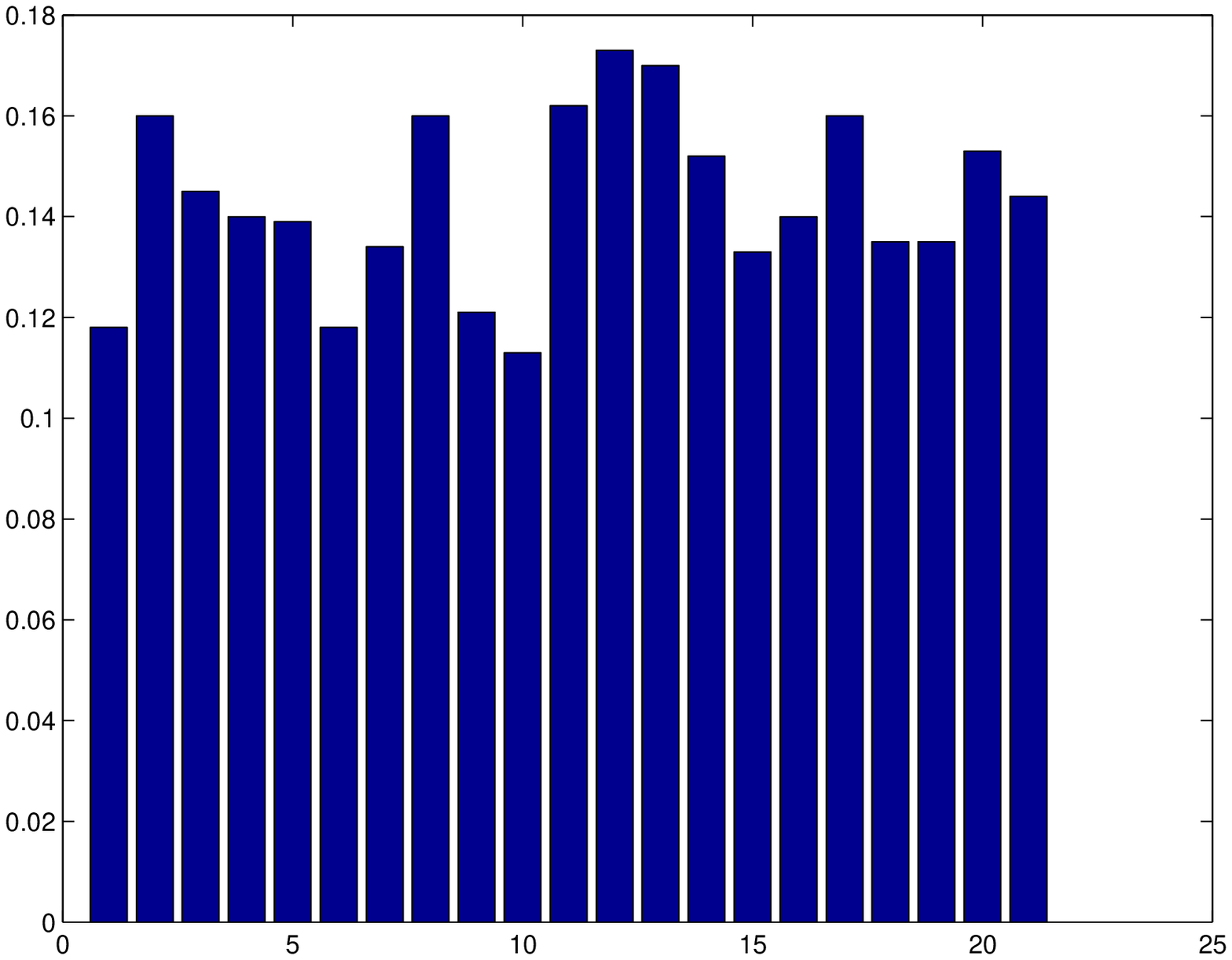}}
    \subfigure[q=4.0]{
    \label{Yeast-infect-local:b} 
    \centering
    \includegraphics[scale=0.25]{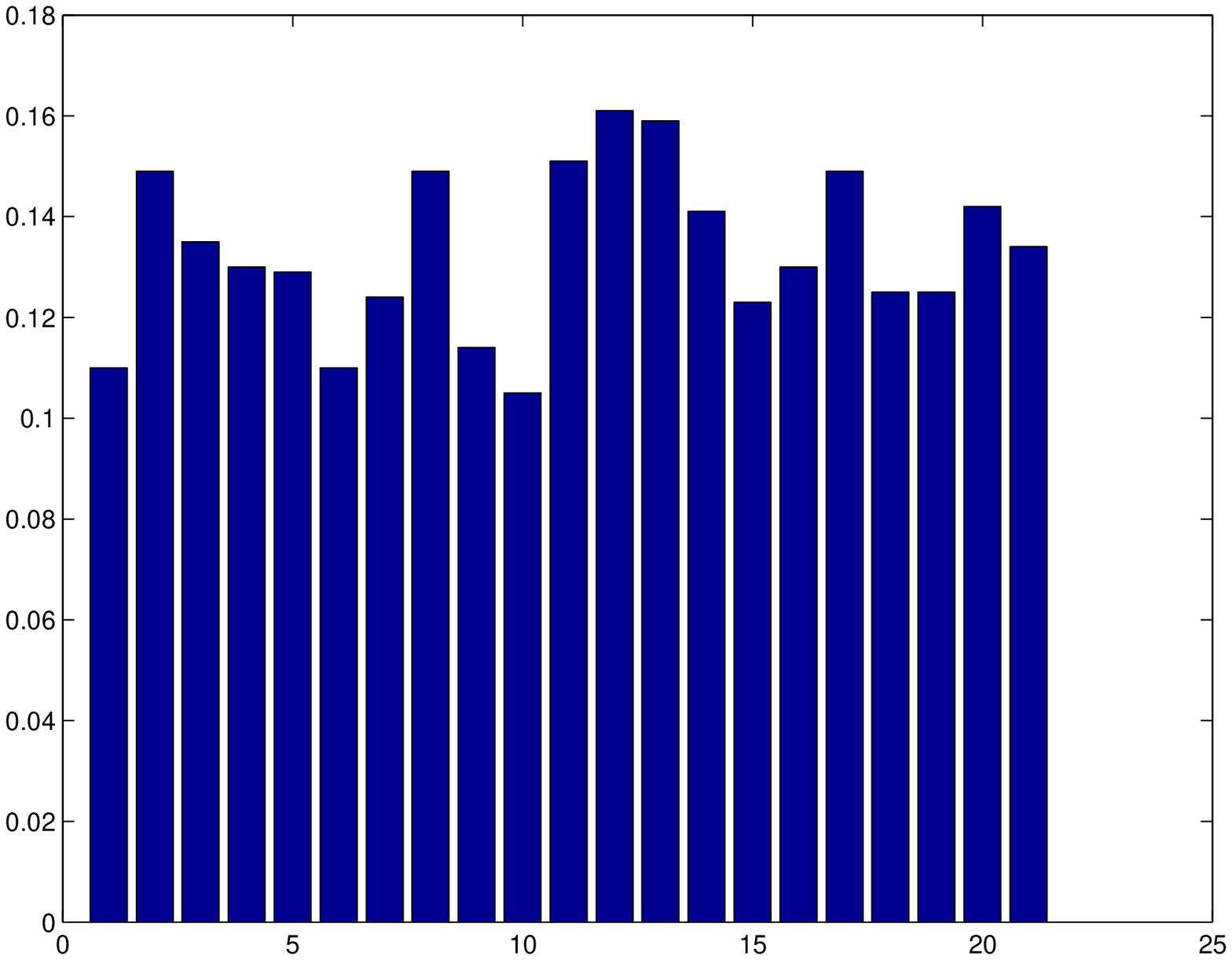}}
    \subfigure[q=4.5]{
    \label{Yeast-infect-local:b} 
    \centering
    \includegraphics[scale=0.25]{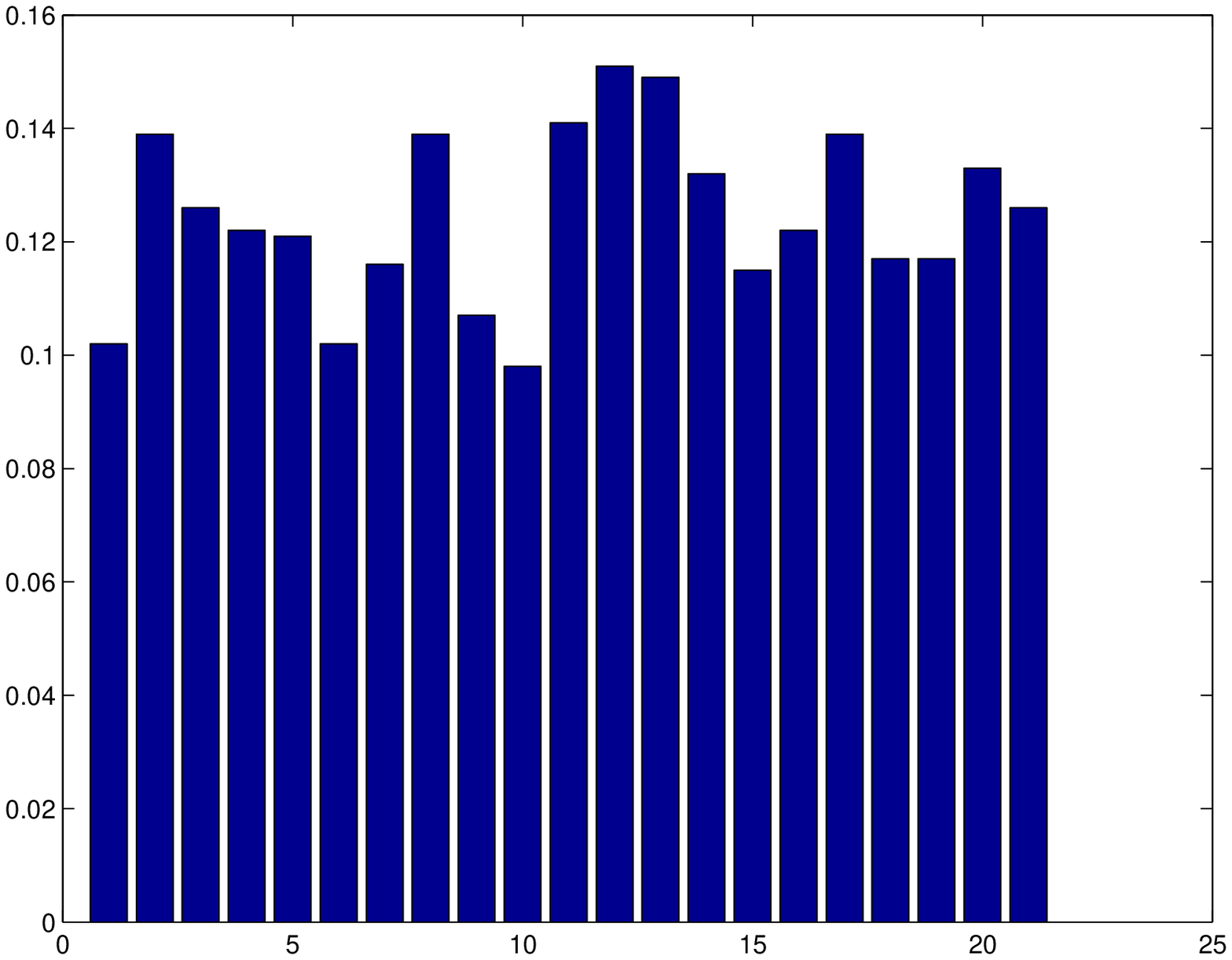}}
    \subfigure[q=5.0]{
    \label{Yeast-infect-local:b} 
    \centering
    \includegraphics[scale=0.25]{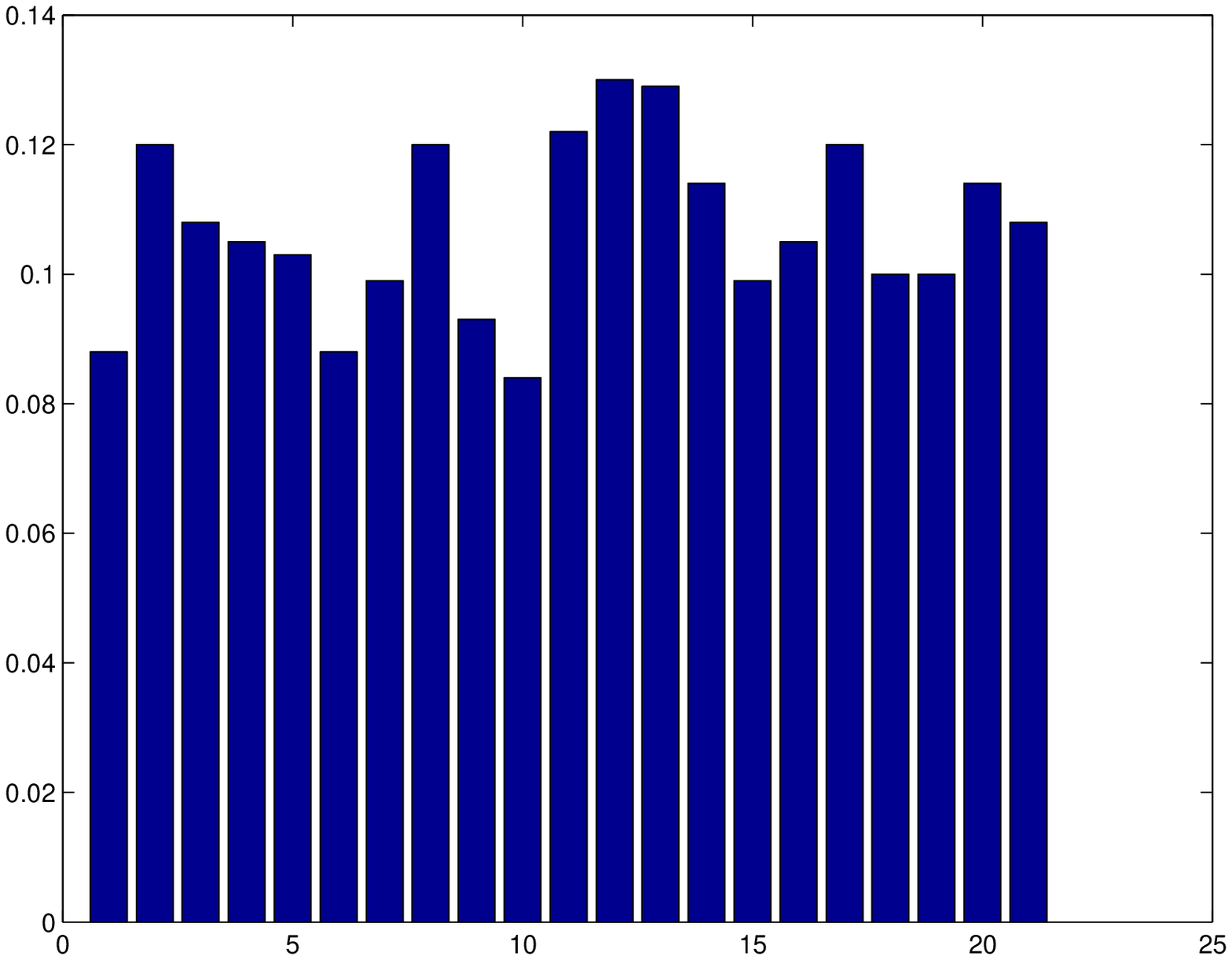}}
    \subfigure[q=5.5]{
    \label{Yeast-infect-local:b} 
    \centering
    \includegraphics[scale=0.25]{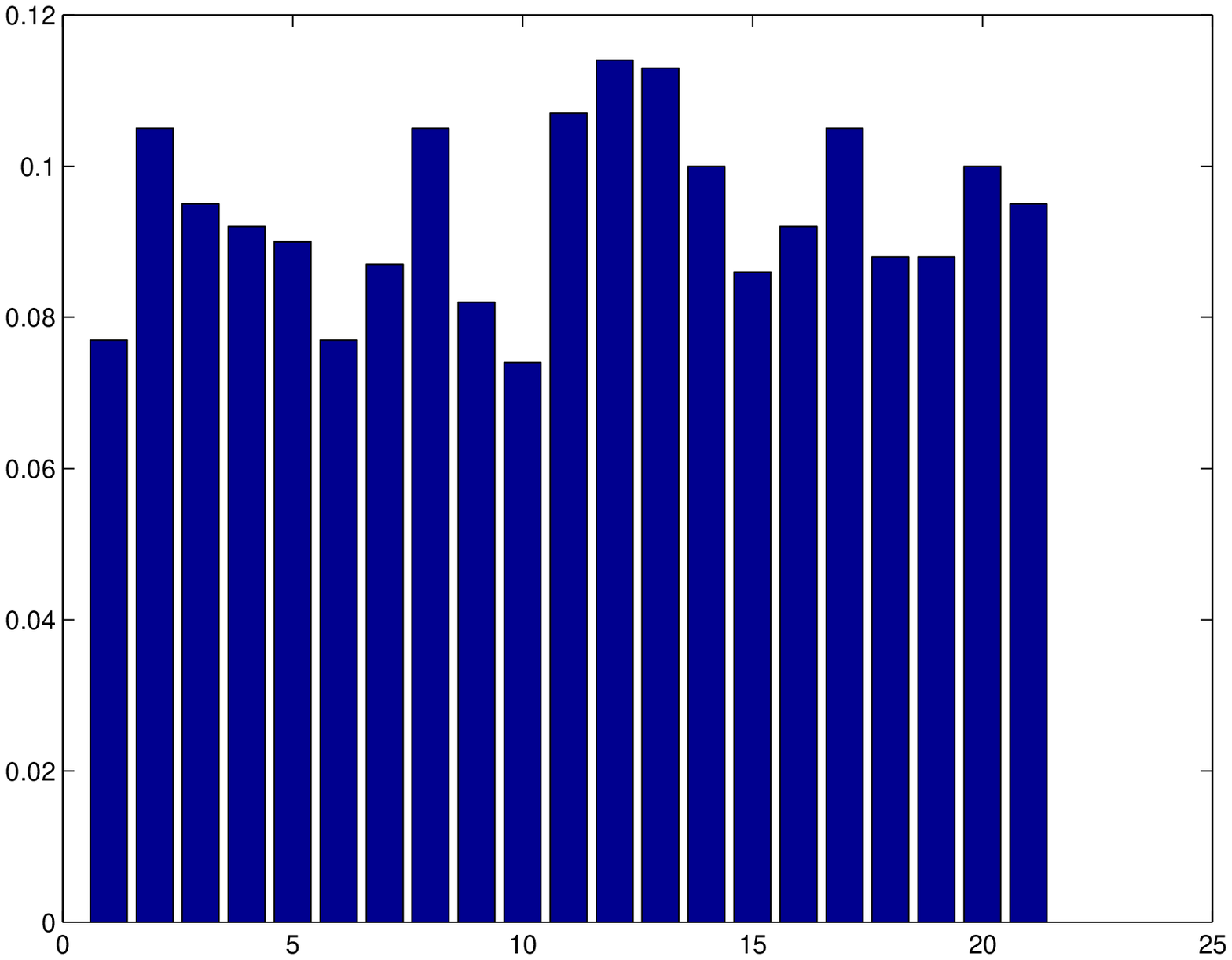}}
    \subfigure[q=6.0]{
    \label{Yeast-infect-local:b} 
    \centering
    \includegraphics[scale=0.25]{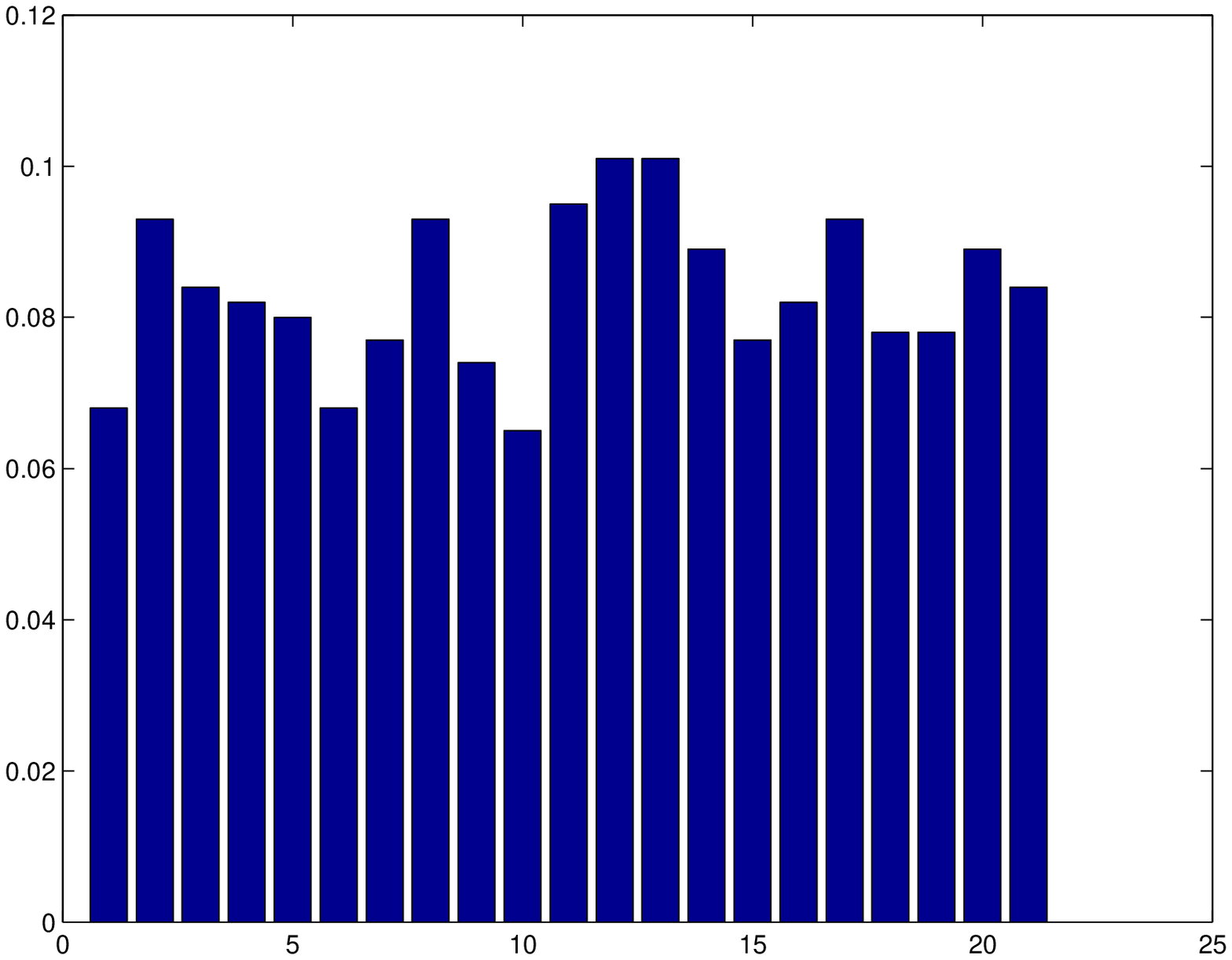}}
    \subfigure[q=6.5]{
    \label{Yeast-infect-local:b} 
    \centering
    \includegraphics[scale=0.25]{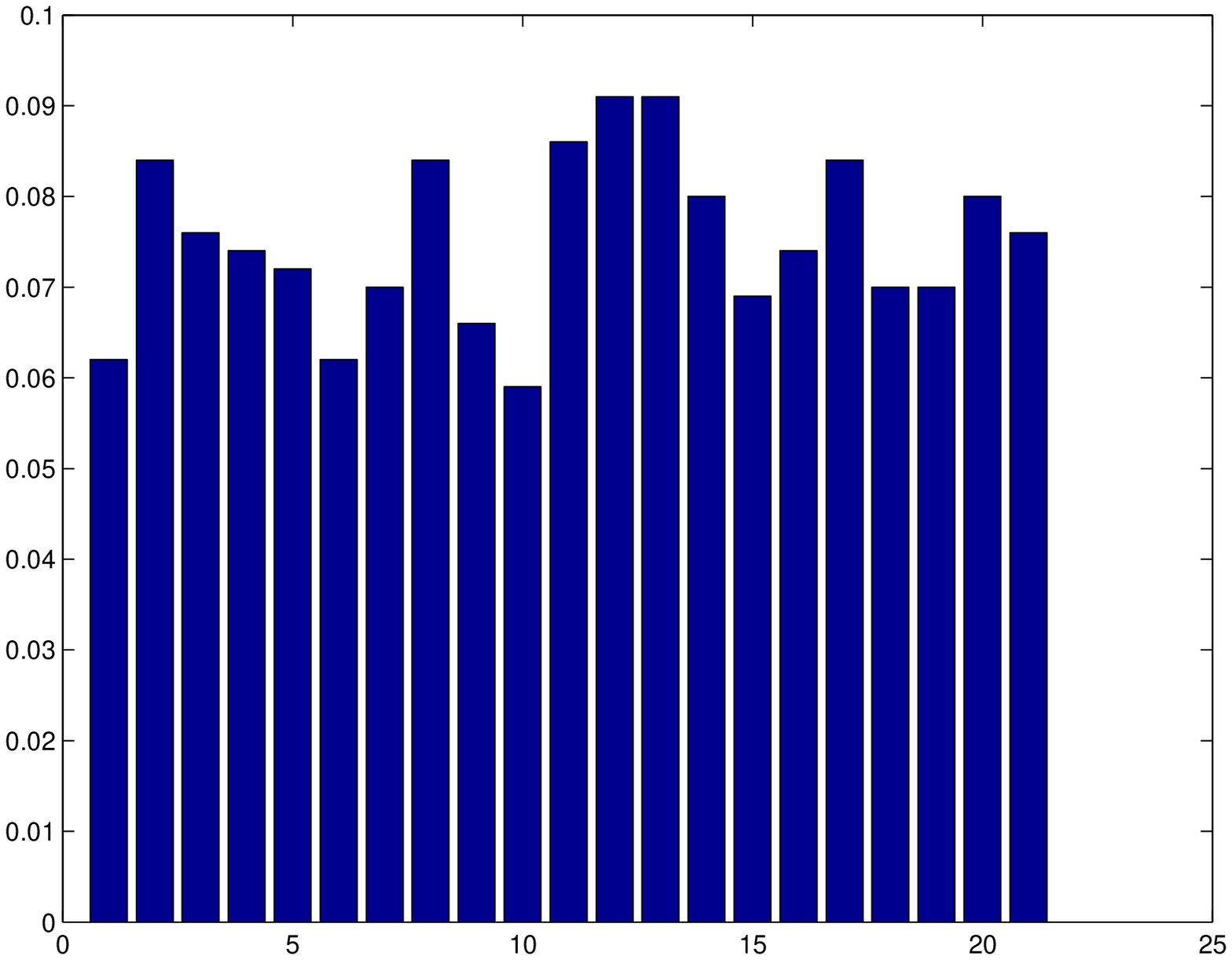}}
    \subfigure[q=7.0]{
    \label{Yeast-infect-local:b} 
    \centering
    \includegraphics[scale=0.25]{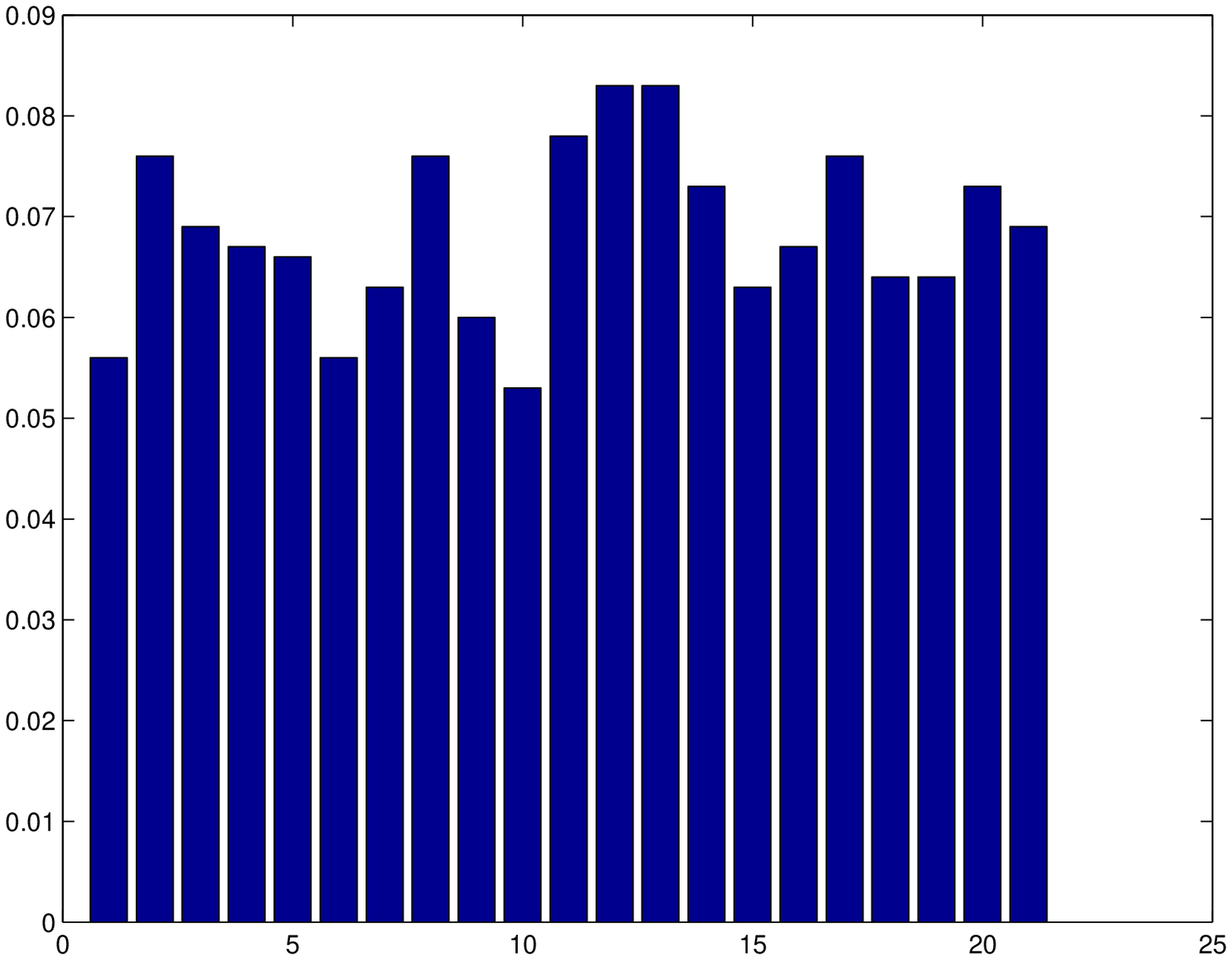}}
  \caption{The figure show the value of the nonextensive local structure entropy of each node in the example network A. The value of $q$ is big than 3.0 and small than 7.0. The caption of the subfigure show the value of $q$. The Abscissa in those subfigure represents the node's number and the ordinate represents the value of nonextensive local structure entropy. }\label{Yeast-infect-local}
\end{figure}

\begin{figure}
    \centering
    \subfigure[q=7.5]{
    \label{Yeast-infect-local:b} 
    \centering
    \includegraphics[scale=0.25]{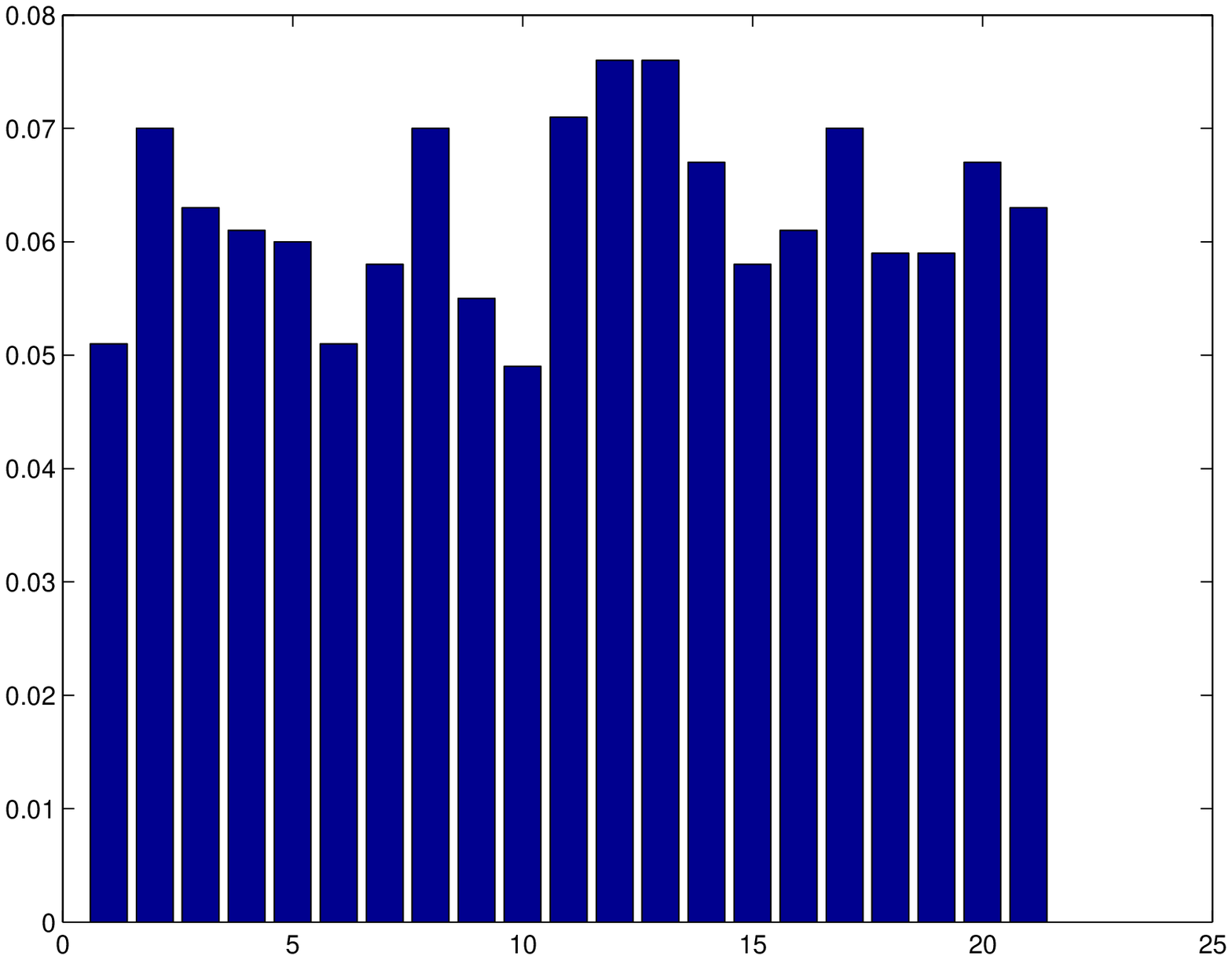}}
    \subfigure[q=8.0]{
    \label{Yeast-infect-local:b} 
    \centering
    \includegraphics[scale=0.25]{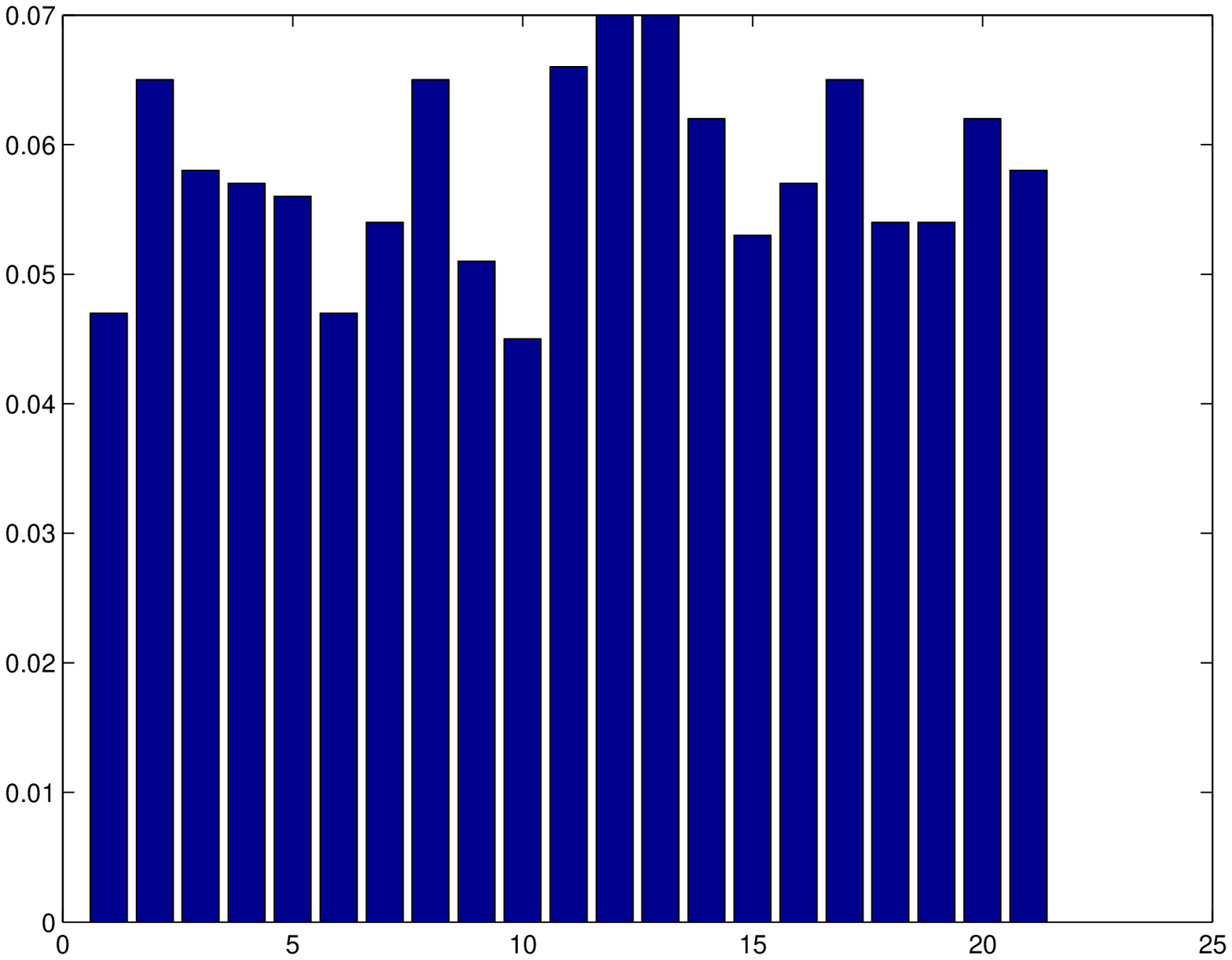}}
    \subfigure[q=8.5]{
    \label{Yeast-infect-local:b} 
    \centering
    \includegraphics[scale=0.25]{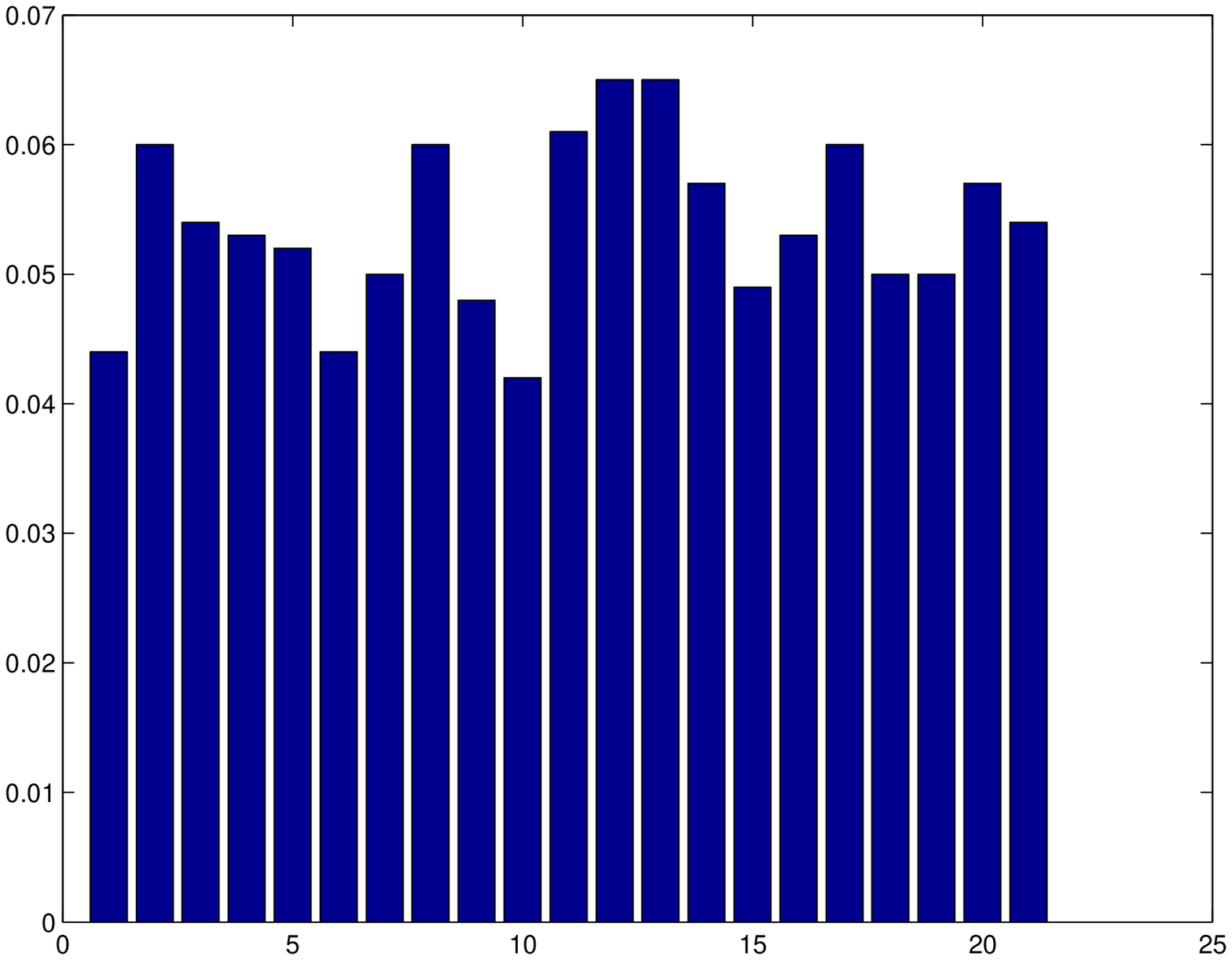}}
    \subfigure[q=9.0]{
    \label{Yeast-infect-local:b} 
    \centering
    \includegraphics[scale=0.25]{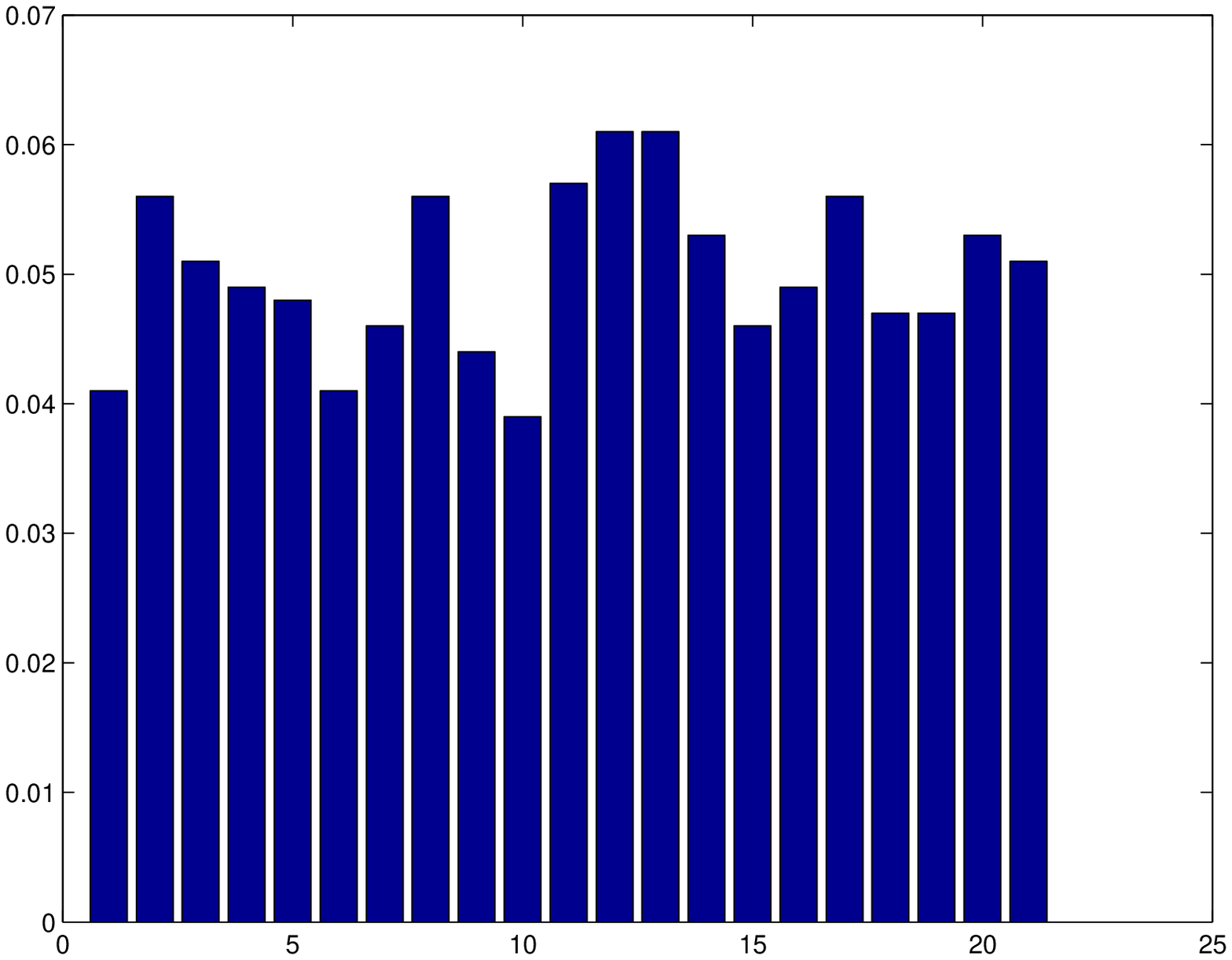}}
    \subfigure[q=9.5]{
    \label{Yeast-infect-local:b} 
    \centering
    \includegraphics[scale=0.25]{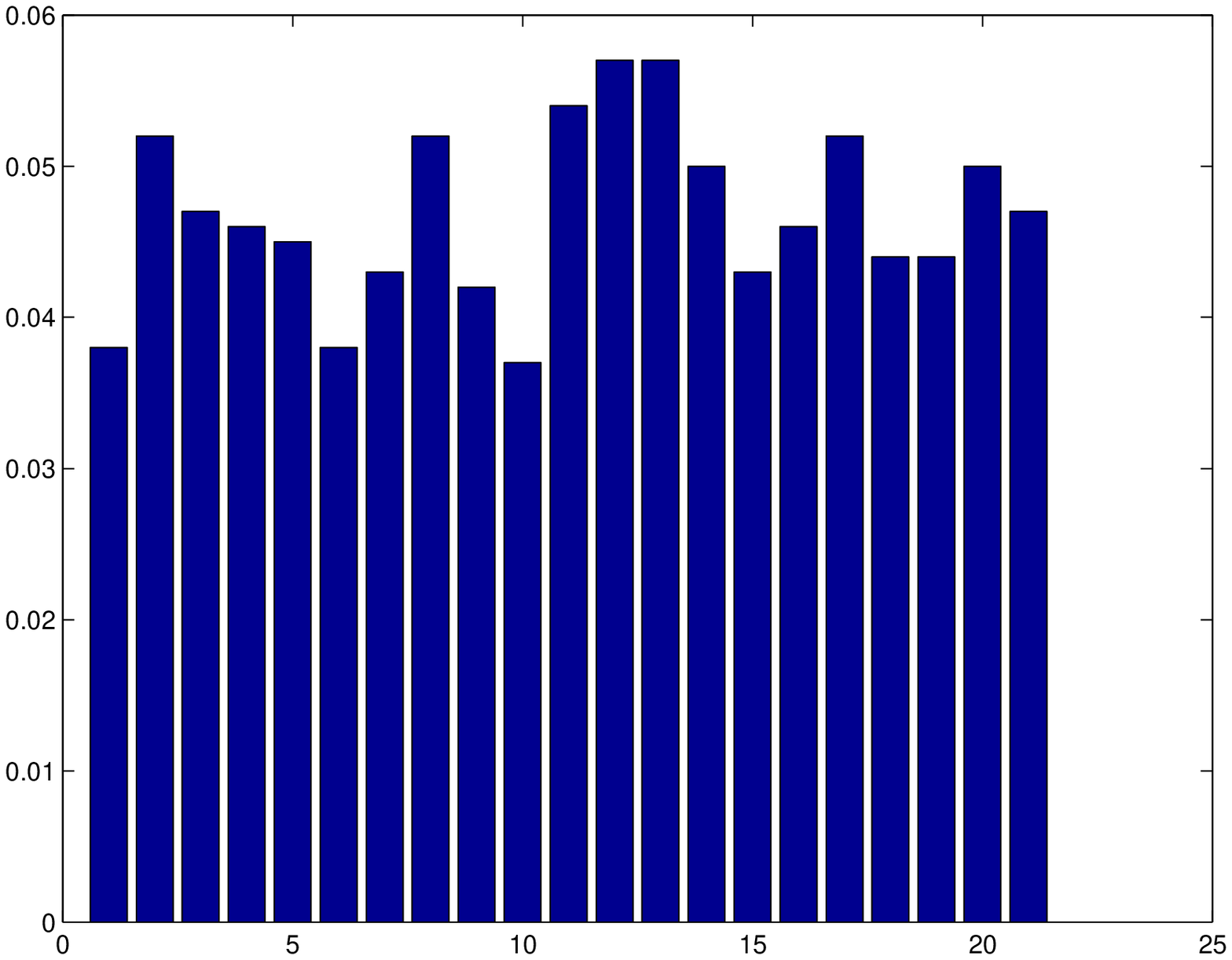}}
    \subfigure[q=10.0]{
    \label{Yeast-infect-local:b} 
    \centering
    \includegraphics[scale=0.25]{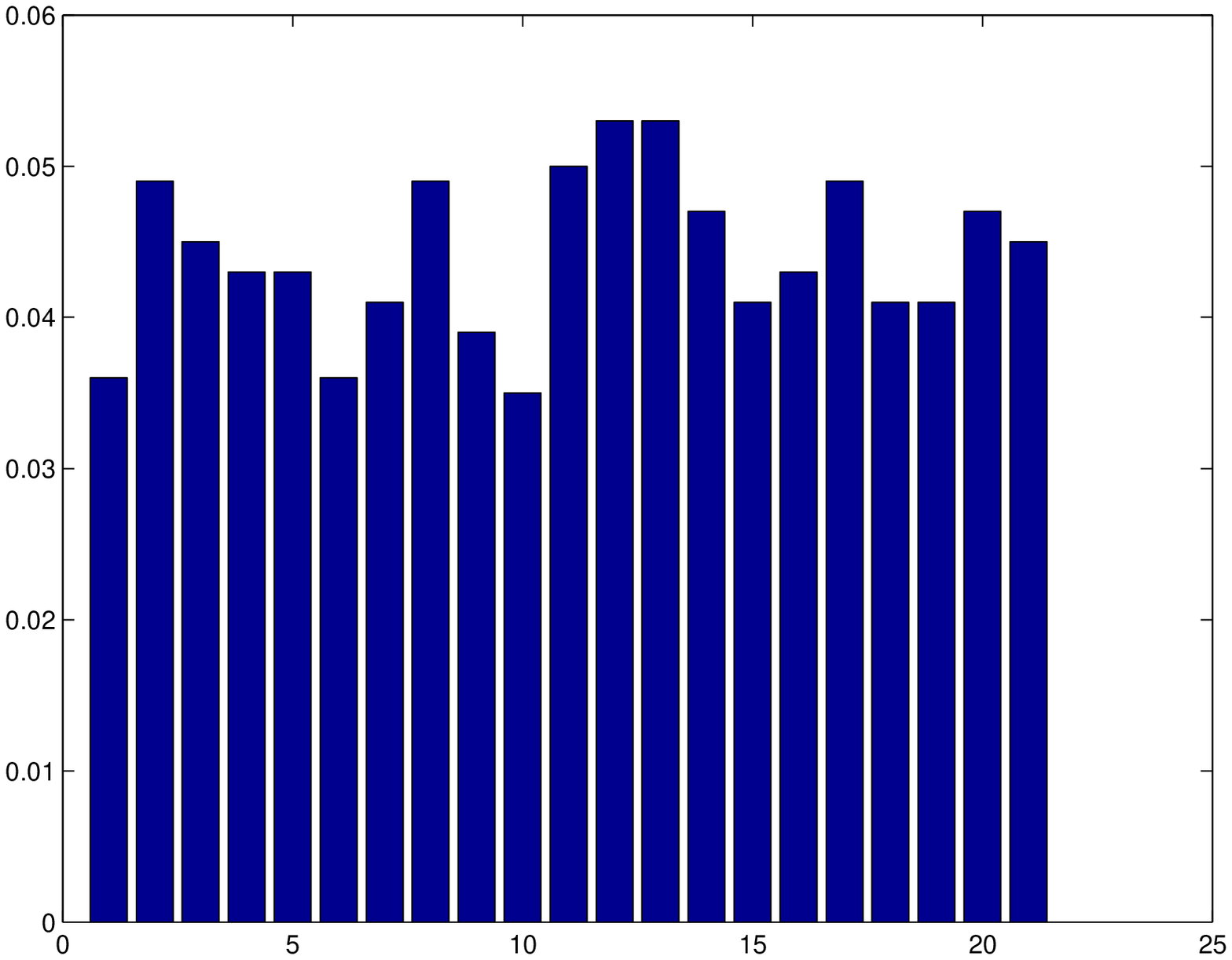}}
  \caption{The figure show the value of the nonextensive local structure entropy of each node in the example network A. The value of $q$ is big than 7.5 and small than 10. The caption of the subfigure show the value of $q$. The Abscissa in those subfigure represents the node's number and the ordinate represents the value of nonextensive local structure entropy. }\label{Yeast-infect-local}
\end{figure}

\begin{table}[htbp]
\tiny
\addtolength{\tabcolsep}{-2pt}
  \centering
  \caption{The order of the influential nodes in the example network A with the change of the value of $q$}
    \begin{tabular}{l|ccccccccccccccccccccc}
    \hline
    Node order&1     & 2     & 3     & 4     & 5     & 6     & 7     & 8     & 9     & 10    & 11    & 12    & 13    & 14    & 15    & 16    & 17    & 18    & 19    & 20    & 21 \\
    \hline
    \rowcolor{gray}
    q=0   & 15    & 7     & 5     & 10    & 19    & 18    & 6     & 1     & 21    & 3     & 20    & 17    & 8     & 2     & 12    & 16    & 4     & 14    & 11    & 13    & 9 \\
    q=0.1 & 15    & 5     & 7     & 10    & 18    & 19    & 1     & 3     & 6     & 20    & 8     & 2     & 21    & 17    & 12    & 16    & 4     & 14    & 11    & 13    & 9 \\
    q=0.2 & 15    & 5     & 7     & 10    & 19    & 18    & 20    & 12    & 8     & 2     & 3     & 17    & 1     & 21    & 6     & 16    & 4     & 14    & 11    & 13    & 9 \\
    q=0.3 & 15    & 5     & 7     & 19    & 18    & 10    & 12    & 8     & 2     & 20    & 17    & 3     & 1     & 21    & 6     & 14    & 16    & 4     & 11    & 13    & 9 \\
    q=0.4 & 15    & 5     & 7     & 19    & 18    & 10    & 12    & 8     & 2     & 20    & 17    & 3     & 21    & 1     & 6     & 14    & 11    & 13    & 16    & 4     & 9 \\
    q=0.5 & 15    & 5     & 7     & 19    & 18    & 10    & 12    & 8     & 2     & 17    & 20    & 3     & 21    & 1     & 6     & 11    & 14    & 13    & 16    & 4     & 9 \\
    q=0.6 & 15    & 5     & 7     & 19    & 18    & 10    & 12    & 8     & 2     & 17    & 20    & 3     & 21    & 1     & 6     & 11    & 13    & 14    & 16    & 4     & 9 \\
    q=0.7 & 15    & 5     & 7     & 19    & 18    & 10    & 12    & 8     & 2     & 17    & 20    & 3     & 21    & 1     & 6     & 13    & 11    & 14    & 16    & 4     & 9 \\
    q=0.8 & 15    & 5     & 7     & 19    & 18    & 12    & 10    & 8     & 2     & 17    & 20    & 3     & 21    & 1     & 6     & 13    & 11    & 14    & 16    & 4     & 9 \\
    q=0.9 & 15    & 5     & 7     & 12    & 19    & 18    & 8     & 2     & 17    & 20    & 10    & 3     & 21    & 1     & 6     & 13    & 11    & 14    & 16    & 4     & 9 \\
    \rowcolor{gray}
    q=1   & 15    & 5     & 7     & 12    & 8     & 2     & 19    & 18    & 17    & 20    & 3     & 10    & 21    & 1     & 13    & 6     & 11    & 14    & 16    & 4     & 9 \\
    q=1.1 & 15    & 5     & 7     & 12    & 8     & 2     & 17    & 19    & 18    & 20    & 3     & 21    & 10    & 1     & 13    & 11    & 14    & 6     & 16    & 4     & 9 \\
    q=1.2 & 15    & 5     & 12    & 7     & 8     & 2     & 17    & 20    & 19    & 18    & 3     & 21    & 10    & 13    & 11    & 1     & 14    & 6     & 16    & 4     & 9 \\
    q=1.3 & 15    & 5     & 12    & 7     & 8     & 2     & 17    & 20    & 19    & 18    & 3     & 21    & 13    & 10    & 11    & 14    & 1     & 16    & 4     & 6     & 9 \\
    q=1.4 & 12    & 15    & 5     & 8     & 2     & 7     & 17    & 20    & 3     & 19    & 18    & 21    & 13    & 11    & 14    & 10    & 1     & 16    & 4     & 6     & 9 \\
    q=1.5 & 12    & 5     & 15    & 8     & 2     & 17    & 7     & 20    & 3     & 19    & 18    & 13    & 21    & 11    & 14    & 10    & 16    & 4     & 1     & 6     & 9 \\
    q=1.6 & 12    & 8     & 2     & 17    & 5     & 15    & 20    & 7     & 3     & 13    & 19    & 18    & 21    & 11    & 14    & 16    & 4     & 10    & 1     & 6     & 9 \\
    q=1.7 & 12    & 8     & 2     & 17    & 5     & 20    & 15    & 7     & 13    & 3     & 11    & 21    & 18    & 19    & 14    & 16    & 4     & 10    & 1     & 6     & 9 \\
    q=1.8 & 12    & 8     & 2     & 17    & 20    & 5     & 13    & 15    & 7     & 3     & 11    & 21    & 18    & 19    & 14    & 16    & 4     & 1     & 10    & 6     & 9 \\
    q=1.9 & 12    & 8     & 2     & 17    & 20    & 13    & 5     & 11    & 15    & 3     & 7     & 21    & 14    & 19    & 18    & 16    & 4     & 1     & 10    & 6     & 9 \\
    q=2   & 12    & 8     & 2     & 17    & 13    & 20    & 11    & 5     & 3     & 15    & 7     & 21    & 14    & 19    & 18    & 16    & 4     & 1     & 10    & 6     & 9 \\
    q=2.2 & 12    & 8     & 2     & 17    & 13    & 20    & 11    & 5     & 3     & 14    & 21    & 15    & 7     & 19    & 18    & 16    & 4     & 1     & 6     & 10    & 9 \\
    q=2.4 & 12    & 13    & 8     & 2     & 17    & 11    & 20    & 14    & 3     & 5     & 21    & 15    & 7     & 18    & 19    & 16    & 4     & 1     & 6     & 10    & 9 \\
    q=2.6 & 12    & 13    & 8     & 2     & 17    & 11    & 20    & 14    & 3     & 21    & 5     & 16    & 4     & 7     & 15    & 19    & 18    & 1     & 6     & 10    & 9 \\
    q=2.8 & 12    & 13    & 2     & 8     & 17    & 11    & 20    & 14    & 3     & 21    & 5     & 16    & 4     & 7     & 19    & 18    & 15    & 1     & 6     & 10    & 9 \\
    q=3   & 12    & 13    & 8     & 2     & 17    & 11    & 20    & 14    & 3     & 21    & 5     & 16    & 4     & 19    & 18    & 7     & 15    & 1     & 6     & 9     & 10 \\
    q=3.2 & 12    & 13    & 11    & 8     & 2     & 17    & 20    & 14    & 3     & 21    & 5     & 16    & 4     & 18    & 19    & 7     & 15    & 1     & 6     & 9     & 10 \\
    q=3.4 & 12    & 13    & 11    & 2     & 8     & 17    & 20    & 14    & 3     & 21    & 16    & 4     & 5     & 19    & 18    & 7     & 15    & 9     & 1     & 6     & 10 \\
    \hline
    \rowcolor{gray}
    q=3.6 & 12    & 13    & 11    & 8     & 2     & 17    & 20    & 14    & 3     & 21    & 16    & 4     & 5     & 19    & 18    & 7     & 15    & 9     & 1     & 6     & 10 \\
    \rowcolor{gray}
    q=3.8 & 12    & 13    & 11    & 8     & 2     & 17    & 20    & 14    & 3     & 21    & 16    & 4     & 5     & 19    & 18    & 7     & 15    & 9     & 1     & 6     & 10 \\
    \rowcolor{gray}
    q=4   & 12    & 13    & 11    & 8     & 2     & 17    & 20    & 14    & 3     & 21    & 16    & 4     & 5     & 19    & 18    & 7     & 15    & 9     & 1     & 6     & 10 \\
    \rowcolor{gray}
    q=4.5 & 12    & 13    & 11    & 8     & 2     & 17    & 20    & 14    & 3     & 21    & 16    & 4     & 5     & 19    & 18    & 7     & 15    & 9     & 1     & 6     & 10 \\
    \rowcolor{gray}
    q=5   & 12    & 13    & 11    & 8     & 2     & 17    & 20    & 14    & 3     & 21    & 16    & 4     & 5     & 18    & 19    & 7     & 15    & 9     & 1     & 6     & 10 \\
    \rowcolor{gray}
    q=5.5 & 12    & 13    & 11    & 8     & 2     & 17    & 20    & 14    & 3     & 21    & 16    & 4     & 5     & 19    & 18    & 7     & 15    & 9     & 1     & 6     & 10 \\
    \rowcolor{gray}
    q=6   & 12    & 13    & 11    & 8     & 2     & 17    & 20    & 14    & 3     & 21    & 16    & 4     & 5     & 19    & 18    & 7     & 15    & 9     & 1     & 6     & 10 \\
    \rowcolor{gray}
    q=6.5 & 12    & 13    & 11    & 8     & 2     & 17    & 20    & 14    & 3     & 21    & 16    & 4     & 5     & 19    & 18    & 7     & 15    & 9     & 1     & 6     & 10 \\
    \rowcolor{gray}
    q=7   & 12    & 13    & 11    & 8     & 2     & 17    & 20    & 14    & 3     & 21    & 16    & 4     & 5     & 19    & 18    & 7     & 15    & 9     & 1     & 6     & 10 \\
    \rowcolor{gray}
    q=7.5 & 12    & 13    & 11    & 8     & 2     & 17    & 20    & 14    & 3     & 21    & 16    & 4     & 5     & 18    & 19    & 7     & 15    & 9     & 1     & 6     & 10 \\
    \rowcolor{gray}
    q=8   & 12    & 13    & 11    & 8     & 2     & 17    & 20    & 14    & 3     & 21    & 16    & 4     & 5     & 18    & 19    & 7     & 15    & 9     & 1     & 6     & 10 \\
    \rowcolor{gray}
    q=8.5 & 12    & 13    & 11    & 8     & 2     & 17    & 20    & 14    & 3     & 21    & 16    & 4     & 5     & 19    & 18    & 7     & 15    & 9     & 1     & 6     & 10 \\
    \rowcolor{gray}
    q=9   & 12    & 13    & 11    & 8     & 2     & 17    & 20    & 14    & 3     & 21    & 16    & 4     & 5     & 19    & 18    & 7     & 15    & 9     & 1     & 6     & 10 \\
    \rowcolor{gray}
    q=9.5 & 12    & 13    & 11    & 8     & 2     & 17    & 20    & 14    & 3     & 21    & 16    & 4     & 5     & 19    & 18    & 7     & 15    & 9     & 1     & 6     & 10 \\
    \rowcolor{gray}
    q=10  & 12    & 13    & 11    & 8     & 2     & 17    & 20    & 14    & 3     & 21    & 16    & 4     & 5     & 19    & 18    & 7     & 15    & 9     & 1     & 6     & 10 \\
    \hline
    \end{tabular}%
  \label{tab:order}%
\end{table}%

The order of the influential nodes in the example network A is shown in the Table \ref{tab:degree}.

\begin{table}[htbp]
\scriptsize
\addtolength{\tabcolsep}{-2pt}
  \centering
  \caption{The degree of each node in the example network A}
    \begin{tabular}{cccccccccccccccccccccc}
    \hline
    Node number & 1     & 2     & 3     & 4     & 5     & 6     & 7     & 8     & 9     & 10    & 11    & 12    & 13    & 14    & 15    & 16    & 17    & 18    & 19    & 20    & 21 \\
    \hline
    Degree & 3     & 3     & 3     & 2     & 5     & 3     & 5     & 3     & 1     & 4     & 2     & 3     & 2     & 2     & 6     & 2     & 3     & 4     & 4     & 3     & 3 \\
    \hline
    \end{tabular}%
  \label{tab:degree}%
\end{table}%

The order of the influential nodes in the example network A is shown in the Table \ref{tab:degree_order}. Where in the Table \ref{tab:degree_order}, the $D_{order}$ represents the order of the influential nodes in the example network A, the $D_{order1}$ represents another order of the influential nodes in the example network A. Both of the $D_{order}$ and the $D_{order1}$ is based on the degree of each node. The $LE_{order}q=x$ represents the order of the influential nodes which is identified by the nonextensive local structure entropy with different value of $q$. The $LE_{order}$ represents the order of the influential nodes which is identified by the local structure entropy.

\begin{table}[htbp]
\scriptsize
\addtolength{\tabcolsep}{-2pt}
  \centering
  \caption{The order of the influential nodes in the example network A}
    \begin{tabular}{l|ccccccccccccccccccccc}
    \toprule
    $D_{order}$ & 15    & 5     & 7     & 10    & 18    & 19    & 1     & 2     & 3     & 6     & 8     & 12    & 17    & 20    & 21    & 4     & 11    & 13    & 14    & 16    & 9 \\
   \rowcolor{gray}
    $D_{order1}$& 15    & 7     & 5     & 10    & 19    & 18    & 6     & 1     & 21    & 3     & 20    & 17    & 8     & 2     & 12    & 16    & 4     & 14    & 11    & 13    & 9 \\
    Degree  & 6     & 5     & 5     & 4     & 4     & 4     & 3     & 3     & 3     & 3     & 3     & 3     & 3     & 3     & 3     & 2     & 2     & 2     & 2     & 2     & 1 \\
     \midrule
     \rowcolor{gray}
    $LE_{order}q=0 $  & 15    & 7     & 5     & 10    & 19    & 18    & 6     & 1     & 21    & 3     & 20    & 17    & 8     & 2     & 12    & 16    & 4     & 14    & 11    & 13    & 9 \\
      \midrule
       $LE_{order}q=1$   & 15    & 5     & 7     & 12    & 8     & 2     & 19    & 18    & 17    & 20    & 3     & 10    & 21    & 1     & 13    & 6     & 11    & 14    & 16    & 4     & 9 \\

        \midrule
    $LE_{order}$ & 15    & 5     & 7     & 12    & 8     & 2     & 19    & 18    & 17    & 20    & 3     & 10    & 21    & 1     & 13    & 6     & 11    & 14    & 16    & 4     & 9 \\
    \bottomrule
    \end{tabular}%
  \label{tab:degree_order}%
\end{table}%

The results in of the test of the local structure entropy based on the nonextensive statistical have show the influence of the nonextensive additivity between each node on the local structure entropy.

When the value of the entropic index $q$ is equal to 0. Then the value of the local structure entropy on each nodes is corresponded to the number of the degree of each node. The influence of the local network is degenerated to the degree's influence on the whole network. Therefore, the order of the influential nodes which is identified by the local structure entropy in the example network A is the same as the the order based on the degree value. In other word, when the value of $q$ is equal to 0, then influence of the components on the local structure entropy is equal to others. The value of the local structure entropy for each node is decided by the node's degree. When the value of $q$ is equal to 1, then the additivity among there components of local structure entropy is based on the degree of the node in the local network. When the value of $q$ is bigger than 3.6, the order of the influential nodes in the example networks is stable. It means that when the value of $q$ is bigger than 3.6, the nonextensive additivity among those components is stable. Change the value of $q$ has no influence on the order of the local structure entropy. The 3.6 is a threshold value of the nonextensive in the local structure entropy of example network A. The $P_{value}$ is used to represents the threshold value of the nonextensive in the local structure entropy.

The details can be illuminated in six parts:

\begin{itemize}
  \item[Case 1] When \textbf{q=0}, the relationship between the components in the local structure entropy is equal to each others. The value of the local structure entropy is decided by the number of the components in it. In the local structure entropy which is based on the degree distribution, the order of the influential nodes in the network is equal to the order which is identified by the degree centrality.
  \item[Case 2] When \textbf{0$<$q$<$1}, these components which have small value are the main part of the local structure entropy.
  \item[Case 3] When \textbf{q=1}, the nonextensive additivity in the local structure entropy is degenerated to the extensive additivity. The nonextensive local structure entropy degenerate to the local structure entropy.
  \item[Case 4] When \textbf{1$<$q$<$$P_{value}$}, the components which have big value are the main part of the local structure entropy.
  \item[Case 5] When \textbf{q=$P_{value}$}, the nonextensive additivity in the local structure entropy achieve to a stable state.
  \item[Case 6] When \textbf{q$>$$P_{value}$}, the order of the influential nodes in the complex networks is achieve to stable. Change the value of $q$ will have no influence on the order of the influential nodes in the complex networks.
\end{itemize}

The order of the influential nodes in the complex can be described in three state: 1. The initial state $Order_{q=0}$. When the value of $q$ is equal to 0, the order of the influential nodes in the complex networks. This state is the same as the order which is identified by the degree centrality. 2. The local structure entropy state $Order_{q=1}$. When the value of $q$ is equal to 1, the order of the influential nodes in the complex networks. The order is decided by the local structure entropy. 3. The stable state $Order_{stable}$. When the value of $q$ is bigger than the nonextensive threshold value, the order of the influential nodes in the complex networks is stable.

The three state of the example network A is shown as follows:

\begin{table}[htbp]
\scriptsize
\addtolength{\tabcolsep}{-2pt}
  \centering
  \caption{The three state of the influential nodes in the example network A}
    \begin{tabular}{l|ccccccccccccccccccccc}
    \toprule
    Node order&1     & 2     & 3     & 4     & 5     & 6     & 7     & 8     & 9     & 10    & 11    & 12    & 13    & 14    & 15    & 16    & 17    & 18    & 19    & 20    & 21 \\
    \midrule
    $Order_{q=0}$ & 15    & 5     & 7     & 10    & 18    & 19    & 1     & 2     & 3     & 6     & 8     & 12    & 17    & 20    & 21    & 4     & 11    & 13    & 14    & 16    & 9 \\
    $Order_{q=1}$  & 15    & 5     & 7     & 12    & 8     & 2     & 19    & 18    & 17    & 20    & 3     & 10    & 21    & 1     & 13    & 6     & 11    & 14    & 16    & 4     & 9 \\
    $Order_{stable}$  & 12    & 13    & 11    & 8     & 2     & 17    & 20    & 14    & 3     & 21    & 16    & 4     & 5     & 19    & 18    & 7     & 15    & 9     & 1     & 6     & 10 \\
    \bottomrule
    \end{tabular}%
  \label{tab:degree_order}%
\end{table}%

The results show that the nonextensive local structure entropy is more useful and more reasonable than the local structure entropy. The nonextensive local structure entropy is a generalised method to identify the influential nodes in the complex networks.

\section{Application}
\label{application}

In this section four real networks is use to analysis the nonextensive in the local structure entropy of it. The four networks are the Zachary's Karate Club network (Karate) \cite{uci}, the US-airport network (Us-airport) \cite{networkdata}, Email networks (Email) \cite{networkdata}and the Germany highway networks (Highway) \cite{nettt}.

The results of the nonextensive threshold value ($P_{value}$) of each network and the three state of the four networks is shown as follows:

\begin{table}[htbp]
\scriptsize
\addtolength{\tabcolsep}{-2pt}
  \centering
  \caption{The three states of the (Karate)network}
    \begin{tabular}{c|cccccccccc|c|cccccccccc}
    \toprule
          & \multicolumn{10}{c}{Top 10 influential nodes}                                                   & \multicolumn{1}{c}{} & \multicolumn{10}{c}{Low 10 influential nodes} \\
    \midrule
    $Order_{q=0}$     & 34    & 1     & 33    & 3     & 2     & 4     & 32    & 24    & 14    & 9     &       & 13    & 18    & 22    & 10    & 15    & 16    & 19    & 21    & 23    & 12 \\
    $Order_{q=1}$     & 14    & 9     & 3     & 32    & 31    & 8     & 4     & 1     & 33    & 2     &       & 6     & 7     & 5     & 11    & 13    & 26    & 27    & 25    & 17    & 12 \\
    $Order_{stable}$ & 12    & 15    & 16    & 19    & 21    & 23    & 20    & 10    & 18    & 22    &       & 3     & 6     & 7     & 2     & 33    & 26    & 25    & 1     & 17    & 34 \\
    \bottomrule
    \end{tabular}%
  \label{tab:Karate}%
\end{table}%

\begin{table}[htbp]
\tiny
\addtolength{\tabcolsep}{-3pt}
  \centering
  \caption{The three states of the (Us-airport) network}
    \begin{tabular}{c|cccccccccc|c|cccccccccc}
    \toprule
          & \multicolumn{10}{c}{Top 10 influential nodes}                                                   & \multicolumn{1}{c}{} & \multicolumn{10}{c}{Low 10 influential nodes} \\
    \midrule
        $Order_{q=0}$   & 118   & 261   & 255   & 152   & 182   & 230   & 166   & 67    & 112   & 201   &       & 277   & 278   & 279   & 280   & 282   & 291   & 294   & 304   & 88    & 114 \\
    \midrule
    $Order_{q=1}$    & 159   & 292   & 172   & 131   & 94    & 109   & 307   & 301   & 310   & 305   &       & 188   & 31    & 22    & 171   & 328   & 330   & 32    & 122   & 332   & 327 \\
    $Order_{stable}$ & 88    & 114   & 241   & 247   & 257   & 268   & 277   & 278   & 279   & 280   &       & 22    & 31    & 13    & 171   & 328   & 330   & 122   & 332   & 32    & 327 \\

    \bottomrule
    \end{tabular}%
  \label{tab:Us-airport}%
\end{table}%

\begin{table}[htbp]
\tiny
\addtolength{\tabcolsep}{-4pt}
  \centering
  \caption{The three states of the (Email) network}
    \begin{tabular}{c|cccccccccc|c|cccccccccc}
    \toprule
          & \multicolumn{10}{c}{Top 10 influential nodes}                                                   & \multicolumn{1}{c}{} & \multicolumn{10}{c}{Low 10 influential nodes} \\
    \midrule
    $Order_{q=0}$     & 105   & 333   & 23    & 16    & 42    & 41    & 196   & 233   & 21    & 76    &       & 791   & 955   & 956   & 959   & 294   & 253   & 261   & 779   & 374   & 644 \\
    \midrule
    $Order_{q=1}$     & 105   & 3     & 39    & 16    & 42    & 54    & 210   & 390   & 50    & 332   &       & 1091  & 1099  & 1118  & 1119  & 1121  & 1130  & 1052  & 1103  & 1125  & 1133 \\
    $Order_{stable}$ & 644   & 236   & 374   & 779   & 247   & 394   & 253   & 261   & 294   & 955   &       & 1091  & 1099  & 1118  & 1119  & 1121  & 1130  & 1052  & 1103  & 1125  & 1133 \\

    \bottomrule
    \end{tabular}%
  \label{tab:Email}%
\end{table}%

\begin{table}[htbp]
\tiny
\addtolength{\tabcolsep}{-4pt}
  \centering
  \caption{The three states of the (Highway) network}
    \begin{tabular}{c|cccccccccc|c|cccccccccc}
    \toprule
          & \multicolumn{10}{c}{Top 10 influential nodes}                                                   & \multicolumn{1}{c}{} & \multicolumn{10}{c}{Low 10 influential nodes} \\
    \midrule
    $Order_{q=0}$    & 693   & 403   & 300   & 410   & 758   & 373   & 217   & 556   & 207   & 331   &       & 247   & 587   & 678   & 898   & 779   & 326   & 1031  & 265   & 787   & 798 \\
    \midrule
    $Order_{q=1}$     & 219   & 393   & 698   & 217   & 404   & 450   & 543   & 267   & 331   & 763   &       & 1154  & 1155  & 1158  & 1160  & 1162  & 1163  & 1165  & 1166  & 1167  & 1168 \\
    $Order_{stable}$ & 265   & 787   & 798   & 326   & 1031  & 779   & 687   & 709   & 161   & 247   &       & 1128  & 1129  & 1131  & 1133  & 1138  & 1144  & 1145  & 1156  & 1159  & 1164 \\

    \bottomrule
    \end{tabular}%
  \label{tab:Highway}%
\end{table}%

\begin{table}[htbp]
  \centering
  \caption{The nonextensive threshold value of the four networks}
    \begin{tabular}{cccc}
    \toprule
    Network &Nodes& edages& $P_{value}$\\
    \midrule
    Karate & 34    & 78    & 4.5 \\
    Us-airport & 332   & 2126  & 7.6 \\
    Email & 1133  & 10902 & 6.2 \\
    Highway  & 1168  & 2486  & 4.1 \\

    \bottomrule

    \end{tabular}%
  \label{tab:threshold value}%
\end{table}%

The threshold value of the nonextensive ($P_{value}$) in the four real networks is shown in the Table \ref{tab:threshold value}. It is clear that the nonextensive in the local structure entropy if not based on the scale of the network.

\section{Conclusion}
\label{conclusion}
The local structure entropy is a new method which is used to identify the influential nodes in the complex networks. In this paper, the local structure entropy of the complex networks is redefined by the nonextensive statistical mechanics. The results of the nonextensive analysis on the local structure entropy shown that when the entropic index $q$ is equal to 0, the order of the influential nodes which is identified by the nonextensvie local structure entropy the same as the degree centrality. When the value of $q$ is equal to 1, then the nonextensive local structure entropy is degenerated to the traditional local structure entropy. When the value of $q$ is bigger than the nonextensive threshold value ($P_{value}$) the order of the influential nodes in the complex networks is stable. 

It is clear that the value of $q$ will influence the property of the local structure entropy, but it also have a range. When the value of $q$ is smaller than $P_{value}$ and bigger than 1, the components with big value in the local structure entropy play an important roles in the local structure entropy. When the value of $q$ is smaller than 1, bigger than 0 the components with small value in the local structure entropy play an important roles in the local structure entropy. When the value of $q$ is equal to 0, then the performance of the local structure entropy is decided by the numbers of the components in the local structure entropy. The nonextensive local structure entropy is degenerates to the degree centrality. 

The new form of the local structure entropy which is defined based on the Tsallis entropy is more reasonable and more useful than the existing one. It is a generalised method which can be used to identify the influential nodes in the complex networks.

\section*{Acknowledgments}
The work is partially supported by National Natural Science Foundation of China (Grant No. 61174022), Specialized Research Fund for the Doctoral Program of Higher Education (Grant No. 20131102130002), R$\&$D Program of China (2012BAH07B01), National High Technology Research and Development Program of China (863 Program) (Grant No. 2013AA013801), the open funding project of State Key Laboratory of Virtual Reality Technology and Systems, Beihang University (Grant No.BUAA-VR-14KF-02). Fundamental Research Funds for the Central Universities No. XDJK2015D009. Chongqing Graduate Student Research Innovation Project (Grant No. CYS14062).

%



\bibliographystyle{elsarticle-num}
\bibliography{zqreference}

\begin{thebibliography}{10}
\expandafter\ifx\csname url\endcsname\relax
  \def\url#1{\texttt{#1}}\fi
\expandafter\ifx\csname urlprefix\endcsname\relax\def\urlprefix{URL }\fi
\expandafter\ifx\csname href\endcsname\relax
  \def\href#1#2{#2} \def\path#1{#1}\fi

\bibitem{albert2000error}
R.~Albert, H.~Jeong, A.~Barab{\'a}si, Error and attack tolerance of complex
  networks, Nature 406~(6794) (2000) 378--382.

\bibitem{newman2003structure}
M.~Newman, The structure and function of complex networks, SIAM Review (2003)
  167--256.

\bibitem{watts1998collective}
D.~Watts, S.~Strogatz, Collective dynamics of ¡®small-world¡¯networks, Nature
  393~(6684) (1998) 440--442.

\bibitem{newman2006structure}
M.~Newman, A.-L. Barab{\'a}si, D.~J. Watts, The structure and dynamics of
  networks, Princeton University Press, 2006.

\bibitem{barthelemy2004betweenness}
M.~Barthelemy, Betweenness centrality in large complex networks, The European
  Physical Journal B-Condensed Matter and Complex Systems 38~(2) (2004)
  163--168.

\bibitem{song2005self}
C.~Song, S.~Havlin, H.~Makse, Self-similarity of complex networks, Nature
  433~(7024) (2005) 392--395.

\bibitem{wei2014informationdimension}
W.~Daijun, W.~Bo, H.~Yong, Z.~Haixin, Y.~Deng, A new information dimension of
  complex networks, Physics Letters A (2014) 1091--1094.

\bibitem{liu2011controllability}
Y.-Y. Liu, J.-J. Slotine, A.-L. Barab{\'a}si, Controllability of complex
  networks, Nature 473~(7346) (2011) 167--173.

\bibitem{arenas2008synchronization}
A.~Arenas, A.~D{\'\i}az-Guilera, J.~Kurths, Y.~Moreno, C.~Zhou, Synchronization
  in complex networks, Physics Reports 469~(3) (2008) 93--153.

\bibitem{barrat2004architecture}
A.~Barrat, M.~Barthelemy, R.~Pastor-Satorras, A.~Vespignani, The architecture
  of complex weighted networks, Proceedings of the National Academy of Sciences
  of the United States of America 101~(11) (2004) 3747--3752.

\bibitem{barabasi2009scale}
A.-L. Barab{\'a}si, et~al., Scale-free networks: a decade and beyond, science
  325~(5939) (2009) 412.

\bibitem{barabasi1999emergence}
A.~Barab{\'a}si, R.~Albert, Emergence of scaling in random networks, Science
  286~(5439) (1999) 509--512.

\bibitem{teixeira2010complex}
G.~Teixeira, M.~Aguiar, C.~Carvalho, D.~Dantas, M.~Cunha, J.~Morais,
  H.~Pereira, J.~Miranda, Complex semantic networks, International Journal of
  Modern Physics C 21~(03) (2010) 333--347.

\bibitem{zhang2014local}
Q.~Zhang, M.~Li, Y.~Du, Y.~Deng, Local structure entropy of complex networks,
  arXiv preprint arXiv:1412.3910.

\bibitem{tsallis1988possible}
C.~Tsallis, Possible generalization of {B}oltzmann-{G}ibbs statistics, Journal
  of Statistical Physics 52~(1-2) (1988) 479--487.

\bibitem{clausius1867mechanical}
R.~Clausius, The mechanical theory of heat: with its applications to the
  steam-engine and to the physical properties of bodies, J. van Voorst, 1867.

\bibitem{shannon2001mathematical}
C.~E. Shannon, A mathematical theory of communication, ACM SIGMOBILE Mobile
  Computing and Communications Review 5~(1) (2001) 3--55.

\bibitem{tsallis2009introduction}
C.~Tsallis, Introduction to Nonextensive Statistical Mechanics, Springer,
  Berlin, 2009.

\bibitem{uci}
Uci network data repository, http://networkdata.ics.uci.edu/data.php?id=105
  (2014).

\bibitem{networkdata}
Pajek datasets, http://vlado.fmf.uni-lj.si/pub/networks/data/ (2014).

\bibitem{nettt}
Tore opsahl, http://toreopsahl.com/datasets/ (2014).

\end{thebibliography}






\end{document}